\documentclass[aps,reprint,twocolumns,longbibliography]{revtex4-1}
\usepackage{graphicx}
\usepackage{color}
\usepackage{upgreek}

\begin{document}

\title{Enhanced force-field calibration via machine learning}
\author{Aykut Argun$^1$, Tobias Thalheim$^2$, Stefano Bo$^3$, Frank Cichos$^2$ and Giovanni Volpe$^{1}$}
\address{$^1$Department of Physics, University of Gothenburg, Origov\"agen 6B, SE-41296 Gothenburg, Sweden,\\
	     $^2$Peter Debye Institute for Soft Matter Physics, Molecular Nanophotonics Group, Leipzig University, Linn\'estra\ss e 5, 0403 Leipzig, Germany, \\
         $^3$ Max Planck Institute for the Physics of Complex Systems, N\"othnitzer Strasse 38, 01187 Dresden, Germany }

\begin{abstract} 
The influence of microscopic force fields on the motion of Brownian particles plays a fundamental role in a broad range of fields, including soft matter, biophysics, and active matter.
Often, the experimental calibration of these force fields relies on the analysis of the trajectories of these Brownian particles.
However, such an analysis is not always straightforward, especially if the underlying force fields are non-conservative or time-varying, driving the system out of thermodynamic equilibrium.
Here, we introduce a toolbox to calibrate microscopic force fields by analyzing the trajectories of a Brownian particle using machine learning, namely recurrent neural networks.
We demonstrate that this machine-learning approach outperforms standard methods when characterizing the force fields generated by harmonic potentials if the available data are limited. 
More importantly, it provides a tool to calibrate force fields in situations for which there are no standard methods, such as non-conservative and time-varying force fields.
In order to make this method readily available for other users, we provide a Python software package named DeepCalib, which can be easily personalized and optimized for specific applications.
\end{abstract}

\maketitle

Measuring microscopic force fields is of fundamental importance to understand microscale systems.
In experimental soft matter, biophysics and active matter, microparticles are often used to probe force fields \cite{jones2015optical, wu2011optoelectronic, braun2013optically, gieseler2020optical}. 
This has been done, for example, to measure the elasticity of cells \cite{mills2004nonlinear, sleep1999elasticity}, inter-particle interactions \cite{su2003interparticle, yada2004direct, paladugu2016nonadditivity}, and non-equilibrium fluctuations \cite{liphardt2002equilibrium, collin2005verification, jun2014high, berut2012experimental}. 
Accurate force calibration is also crucial to study molecular motors \cite{toyabe2010nonequilibrium} and microscopic heat engines \cite{blickle2012realization, quinto2014microscopic, martinez2016brownian, martinez2017colloidal, argun2017experimental, schmidt2018microscopic}.
Sometimes the calibration of the force field needs to be even done in real time \cite{gavrilov2014real}.
Disentangling the deterministic force fields from the unavoidable Brownian noise in these systems requires care and has a direct impact on the quality of the experimental results.

Having access to an arbitrarily large amount of data, the profile of a generic force field can be directly estimated by averaging the particle displacements at different positions and times (see, e.g., \cite{wu2009direct, friedrich2011approaching}). 
However, there are many experimental situations where this is not feasible.
As a consequence, several methods have been developed for the most common force fields, which have become standard in various fields \cite{jones2015optical, ciliberto2017experiments, gieseler2020optical}. 

A particularly well-studied case is that of the force field $F_{\rm h}(x) = -k x$ (where $k$ is the stiffness and $x$ the particle position with respect to the equilibrium) generated by a harmonic potential $U_{\rm h}(x) = {1\over2}k x^2$. 
This case is particularly interesting because it approximates the force field near any stable equilibrium, such as that experienced by microscopic particles held in optical, magnetic or acoustic traps \cite{jones2015optical, gieseler2020optical}.
The simplest approach to its calibration exploits the relation between the experimental probability distribution $\rho(x)$ and the potential, i.e., $U_{\rm h}(x) = -k_{\rm B} T \ln N \rho(x)$, 
where $N$ is the normalization factor, $k_{\rm B}$ is the Boltzmann constant and $T$ is the absolute temperature, from which the force can be derived as $F_{\rm h}(x) = - {\partial \over \partial x} U_{\rm h}(x)$. 
Several additional methods are also available. These methods use the temporal information contained in the particle trajectory, extracted by calculating the autocorrelation function \cite{jones2015optical, gieseler2020optical}, the power spectral density \cite{berg2004power}, or the recently developed algorithm FORMA, a maximum likelihood estimator  based on linear regression \cite{garcia2018high}.
All these methods work well with long trajectories, while their performance declines when only short trajectories are available.

Some of the methods used for the calibration of harmonic traps can be generalized to more complex force fields. For example, the potential method can, in principle, be used to characterize any conservative force field at thermodynamic equilibrium; however, the required amount of data grows exponentially for complex potential landscapes because the probe particle must be given enough time to explore the entire configuration space.
Standard methods for the calibration of even more complex force fields, such as non-conservative or time-varying force fields, are not readily available.
The calibration becomes particularly complex when dealing with a limited amount of data, such as when real-time calibration is necessary \cite{jun2014high}.
In fact, developing methods for the calibration of some specific examples of these force fields is a very active field of research \cite{bottcher2006reconstruction, turkcan2012bayesian, bera2017fast, ciliberto2017experiments, garcia2018high, frishman2020learning}.

In this article, we demonstrate numerically and experimentally that machine learning can  efficiently calibrate the force field experienced by a Brownian particle. 
Specifically, we employ a recurrent neural network (RNN)  \cite{lipton2015critical} because RNNs have been proven very successful at tasks requiring the analysis of time series, such as natural language recognition and translation \cite{graves2013hybrid, wu2016google, han2017ese}, event prediction \cite{gers1999learning}, and anomalous diffusion characterization \cite{bo2019measurement}.
We demonstrate that this RNN-powered method outperforms standard calibration techniques when calibrating a harmonic potential using only a short trajectory.
Then, we demonstrate that it can also be used to calibrate force fields for which standard calibration techniques do not exist, namely bistable, non-conservative, and time-varying force fields.
In order to make this approach readily available for other users, we provide a Python software package, called DeepCalib \cite{DC}, which can be readily personalized and optimized for the needs of specific users and applications.

\section{Results}

Machine-learning-powered techniques have been particularly successful in data analysis,
emerging as an ideal method to study systems for which only limited data is available or no standard approaches are available \cite{zdeborova2017machine, cichos2020machine}. 
In particular, artificial neural networks \cite{nielsen2015neural, chollet2018keras} provide a powerful way to automatically extract information from data. 
They belong to the class of supervised machine-learning methods. 
Unlike standard algorithmic approaches that use explicit mathematical recipes in order to obtain the sought-after results,  supervised machine-learning methods are trained with large data sets associated with the corresponding ground truth in order to determine the optimal processing to estimate this ground truth from the input data.
The learning task is typically a classification (where the ground truth indicates to which class the input belongs, e.g., determining if an image contains a cat or a dog) or a regression (where the ground truth is the numerical value of a quantity, e.g., inferring a parameter from a physical experiment). 

Neural networks are composed of artificial neurons connected by adjustable weights. 
These neurons are often arranged in layers.
The neurons in a layer perform a nonlinear transformation of the inputs they receive and feed their results to the neurons of the subsequent layer.
The final layer returns an estimate of the ground truth corresponding to the original input. 
The training process consists of iteratively adjusting the weights of the neural network in order to decrease the distance between the output and the ground truth of the sample so that the network progressively learns to associate the input data to the correct ground truth. 
This is usually achieved by back-propagating the estimation error through the layers \cite{mcclelland1986parallel}. Once the neural network is trained, it can be used to predict the features of data it has never seen before. 

Neural networks have recently been shown by physicists to be a powerful tool for classification and parametrization of stochastic phenomena, e.g., to determine anomalous diffusion exponents \cite{bo2019measurement, granik2019single} (also recently done using random forests \cite{munoz2020single}), the arrow of time \cite{seif2019machine}, and the position of particles \cite{hannel2018machine, helgadottir2019digital}, as well as in microscopy \cite{barbastathis2019use} and simulations of hydrodynamic interactions \cite{gibson2019machine}, and optical forces \cite{lenton2020machine}.

This success of neural networks in analyzing experimental data, motivated us to test their performance in reconstructing microscopic force fields. 
This has led us to develop a force-field calibration method based on the use of RNNs, which are especially well-suited to handle time series  because they process the input data sequence iteratively and, therefore, explicitly model the sequentiality of the input data.
We name this method and the corresponding software package DeepCalib \cite{DC}.
Given a force field characterized by a set of parameters (e.g., a harmonic force field characterized by its stiffness $k$), we train the RNN to infer these parameters from short trajectories of Brownian particles moving in such force fields.
Specifically, DeepCalib analyzes an input trajectory (typically corresponding to 1000 time steps) using an RNN with 3 long short-term memory (LSTM) layers (with 1000, 250 and 50 nodes, respectively) and outputs the estimated values of the force-field parameters.
We choose LSTMs because their architecture manages to retain a combination of short time as well as longer term correlations without making the training procedure excessively unstable \cite{lipton2015critical}.
Further, LSTMs have been shown to perform well on short stochastic time series \cite{bo2019measurement}.
For the training of the RNN, we use simulated trajectories, for which we know the ground-truth values of the force-field parameters, to iteratively adjust the weights in the nodes in the LSTM layers using the back-propagation training algorithm \cite{mcclelland1986parallel}.
Finally, we test the performance of the trained RNN on experimental trajectories of Brownian particles in force fields that we generate using a thermophoretic feedback trap \cite{braun2015single}.

In the following sections, we demonstrate that DeepCalib can be used to estimate a large variety of force fields from stochastic trajectories. 
We start by considering the paradigmatic case of a harmonic trap, showing that DeepCalib outperforms standard techniques for short trajectories.
Then, we move to more complicated scenarios: a double-well potential, a non-conservative force field, and a time-varying force field for which no simple general calibration method exists. 
We provide the source code of DeepCalib together with example files that reproduce all presented results \cite{DC}. 
This code can be easily adapted and optimized for the needs of specific users and applications.

\subsection{Harmonic potential}

%%%%%%%%%%%%%%%%%%%%%%%%%%%%%
%% FIGURE 1
%%%%%%%%%%%%%%%%%%%%%%%%%%%%%
\begin{figure}[b]
	\includegraphics[width=.45 \textwidth]{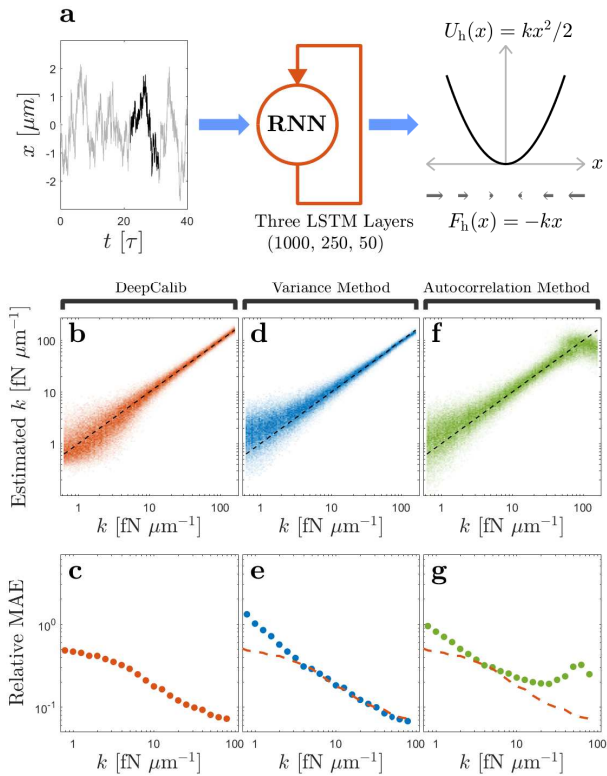}
	\caption{
		{\bf Calibration of harmonic potentials (simulations).} 
		{\bf a}, Simulated trajectory (gray line) of a Brownian particle in a harmonic trap. DeepCalib employs a recurrent neural network (RNN, 3 LSTM layers of sizes 1000, 250 and 50) to estimate the stiffness parameter ($k$) from a short section of the trajectory (black line). 
		{\bf b}, Distribution (orange density plot) of $k$ estimated by DeepCalib using simulated trajectories; the ground truth value of $k$ is provided by the black dashed line. 
		{\bf c}, Relative mean absolute error (MAE) (orange dots) of the DeepCalib estimations as a function of $k$. 
		{\bf d}, Distribution (blue density plot) and, {\bf e}, relative MAE (blue dots) of $k$ estimated by the variance method. 
		{\bf f}, Distribution (green density plot) and, {\bf g}, relative MAE (green dots) of $k$ estimated by the autocorrelation method.
		The dashed orange lines in {\bf e} and {\bf g} reproduce the relative MAE for DeepCalib from {\bf c} for ease of comparison: DeepCalib provides smaller errors for the entire range of $k$. 
		These results are obtained from a test dataset of $10^5$ trajectories, each sampled 1000 times every $10\,{\rm ms}$. Both training and test trajectories are generated with $k$ values uniformly distributed in logarithmic scale.
		See also Example 1a of the DeepCalib software package \cite{DC}.
	}
	\label{fig1}
\end{figure}
%%%%%%%%%%%%%%%%%%%%%%%%%%%%%

In order to benchmark the performance of DeepCalib, we start by considering the simple case of a force field generated by a harmonic potential, for which many efficient standard calibration methods already exist.
Harmonic traps are widely studied because they represent good approximations to more complex profiles near their stable equilibria, and they are easy to realize experimentally and to analyze.  
A Brownian particle in a harmonic trap, in the overdamped limit, is described by the Langevin equation \cite{volpe2013simulation}:
\begin{equation}
	{{\rm d} x \over {\rm d} t} 
	= 
	-{k \over \gamma} x 
	+ 
	\sqrt{2k_{{\rm B}} T \over \gamma}
	\xi(t),
\end{equation}
where $\gamma$ is the friction coefficient and $\xi(t)$ is uncorrelated Gaussian noise with unitary variance. 
An example of a simulated trajectory is shown in Fig.~\ref{fig1}a.
To calibrate this force field, one needs to estimate the stiffness $k$.

We train the RNN using simulated trajectories with different $k$.
The friction coefficient $\gamma$ is randomly varied by 5\% around its nominal value in order for the RNN to gain tolerance against small fluctuations in the friction. 
Since we want to train the RNN to estimate accurately stiffness values that can vary over a few orders of magnitude (from $1$ to $100\,{\rm fN \upmu m^{-1}}$), we draw the values of $k$ from a distribution that is uniform in logarithmic scale (from $10^{-0.5}$ to $10^{3.5}\,{\rm fN \upmu m^{-1}}$).
This is a challenging task because the range of $k$ is very broad and the trajectory is very short (an example is the black portion of the trajectory in Fig.~\ref{fig1}a).
Importantly, the training range of $k$ is wider than the desired measurement range in order to ensure that the RNN is properly trained also for the expected edge cases.
Overall, we train the RNN using $10^7$ trajectories corresponding to $10\,{\rm s}$ and sampled 1000 times (time step $10\,{\rm ms}$).
We continuously generate new trajectories (so that the RNN is never trained twice with the same trajectory, avoiding any risk of overtraining) and split them in batches of increasing size (from 32 to 2048, so that, at the beginning, the RNN optimization process can freely explore a large parameter space and, gradually, it gets progressively annealed towards an optimal parameter set \cite{smith2017don}).
The training process is efficient and takes about four hours on a GPU-enhanced laptop (Intel Core i7 8750H, Nvidia GeForce GTX 1060). 
For further details on the model and the training, see also Example 1a of the DeepCalib software package \cite{DC}.  

The estimations done by DeepCalib are shown in Fig.~\ref{fig1}b (orange distribution) in comparison with the ground truth (black dashed line), while the corresponding relative mean absolute error (MAE) is shown in Fig.~\ref{fig1}c (orange dots). 
DeepCalib provides accurate results for the entire range of $k$, significantly improving  its performance at larger $k$. 
This is expected, because the fluctuations of the particle position in the trap are inversely proportional to $k$, so that for larger $k$ the 10-s trajectory is able to explore the trapping potential more efficiently.

We now compare DeepCalib to two of the
most commonly used methods for estimating the stiffness of a harmonic trap: the variance method and the autocorrelation method \cite{jones2015optical, gieseler2020optical}. 
The variance method (Figs.~\ref{fig1}d--e) determines $k$ from the measurement of the variance of the particle position of the trap:
\begin{equation}
	k 
	=
	\frac{
		k_{\rm B} T 
	}{ 
		\langle x^2  \rangle
	}.
\end{equation}
The autocorrelation method (Figs.~\ref{fig1}f--g) determines $k$ by fitting the decorrelation curve of the particle position in the trap: 
\begin{equation}
	\langle x(t+\Delta t) x(t) \rangle 
	= 
	\frac{
		k_{\rm B} T
	}{
		k
	} 
	e^{ - \Delta t / \tau },
\end{equation}
where $\tau = \gamma/k$ is the characteristic time of the trap.
In both cases, $\langle \cdot \rangle$ represents averaging over time.

The estimations of $k$ obtained with the variance and autocorrelation methods present the distributions shown in Fig.~\ref{fig1}d (blue density plot) and in Fig.~\ref{fig1}f (green density plot), respectively. 
The autocorrelation method provides slightly more accurate results than the variance method when $k$ is small; however, it becomes less accurate when $k$ is large because individual data samples in the trajectory become excessively uncorrelated. 
The corresponding relative MAE of the variance and autocorrelation methods are shown in Fig.~\ref{fig1}e (blue dots) and in Fig.~\ref{fig1}g (green dots), together with the comparison with DeepCalib's performance (orange dashed line).
DeepCalib outperforms the other methods over the whole range of $k$, with the difference being more pronounced for smaller $k$ values, where the measurement is more challenging.

\subsection{Experimental setup and initial experimental validation}

%%%%%%%%%%%%%%%%%%%%%%%%%%%%%
%% FIGURE 2
%%%%%%%%%%%%%%%%%%%%%%%%%%%%%
\begin{figure} [t]
	\includegraphics[width=.45 \textwidth]{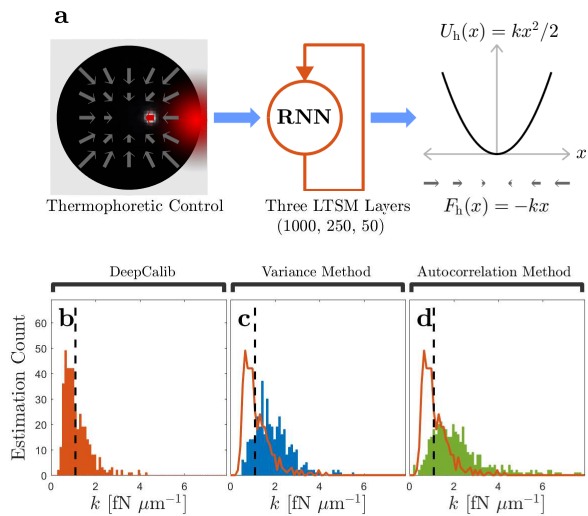}
	\caption{
		{\bf Calibration of harmonic potentials (experiments).} 
		{\bf a}, A fluorescent particle ($R = 100$ nm) is trapped in a harmonic potential using a thermophoretic feedback trap. 
		By focusing a laser on a nanofabricated chrome film (gray structure) with a disk-shaped hole (black region, diameter $15\,{\rm \upmu  m}$), one generates a thermal gradient across the particle and, therefore, a thermophoretic force pushing the particle towards the center of the ring. By adjusting the laser position and intensity as a function of the measured position of the particle, it is possible to control the direction and strength of the thermophoretic force.
		The resulting particle trajectories are measured by digital video microscopy and used by the RNN in order to estimate the trap stiffness $k$. 
		{\bf b}-{\bf d}, Distribution of the $k$ estimated by DeepCalib (orange histogram in {\bf b}, analyzed using the same RNN as in Fig.~\ref{fig1}), the variance method (blue histogram in {\bf c}), and the autocorrelation method (green histogram in {\bf d}) for 400 10-s segments of a single 500-s trajectory (each segment corresponds to 1000 samples taken every $10\,{\rm ms}$).
		The black dashed vertical line represents the estimation of $k$ using the full length of the trajectory using potential method, which we take as ground truth.
		The orange lines in {\bf c} and {\bf d} reproduce the histogram for DeepCalib from {\bf b} for ease of comparison: DeepCalib outperforms the other methods featuring lower variance and lower bias. 
		See also Example 1b of the DeepCalib software package \cite{DC}.
	}
	\label{fig2}
\end{figure}
%%%%%%%%%%%%%%%%%%%%%%%%%%%%%

So far, we have demonstrated how DeepCalib performs on simulated test data that are obtained similarly to the training data set. 
In order to test DeepCalib in a realistic situation, we now investigate the performance of the same RNN discussed in the previous section, trained on simulated data (Fig.~\ref{fig1}), on experimental trajectories. 

The experimental setup to obtain the trajectories consists of a feedback trapping system that enables to generate a wide variety of force fields \cite{braun2015single, franzl2019thermophoretic}.
We measure the Brownian motion of a single 200-nm-diameter polystyrene particle (ThermoFisher Scientific, F8810) in an aqueous environment, confined by dynamic temperature fields to a circular region of a UV-lithographically-fabricated nanostructure (Fig.~\ref{fig2}a) \cite{franzl2019thermophoretic}. 
The confinement in these temperature fields occurs as a result of thermophoretic drifts of the particle due to temperature-dependent solute--solvent interactions \cite{wurger2010thermal, bregulla2016thermo} (red arrow, Fig.~\ref{fig2}a). 
The microscopic origin of these drifts is manifold and summarized in the thermodiffusion coefficient $D_{\rm T}$. 
As it is usually the case, also in our experiment, $D_{\rm T}$ has a positive sign, which means that the corresponding objects move towards colder regions in the temperature landscape. 
A thermophoretic drift velocity $\vec{v}_{\rm T}=-D_{\rm T}\nabla T$ can be assigned to this directed motion, which is proportional to the temperature gradient $\nabla T$. 
The relative strength of the thermophoretic motion of particles in liquids is given by the ratio $S_{\rm T}=D_{\rm T}/D$, which is also known as Soret coefficient. 
Typical values for the Soret coefficient are in the range of $0.01-10\,{\rm K^{-1}}$ \cite{wurger2010thermal}. 
In the thermophoretic trapping setup, temperature gradients are generated by the conversion of optical energy of a focused 808-nm laser beam (Pegasus Lasersysteme, PL.MI.808.300, beam waist $\omega_0 \approx 500\,{\rm nm}$, Fig.~\ref{fig2}a) positioned on the circumference of a circular hole with diameter of $15\,{\rm \upmu  m}$ in an otherwise continuous chrome film  (thickness $30\,{\rm nm}$). 
Temperature differences between the rim and and the trapping center are typically on the order of $\Delta T \approx 10\,{\rm K}$.
The current position of the particle is obtained by the fluorescence emitted by the particle under homogeneous illumination with an excitation laser ($\lambda=532\,{\rm nm}$, Pusch OptoTech), is recorded via an EMCCD camera (Andor iXon 3) at a frequency of $100\,{\rm Hz}$, and is evaluated in real time using a custom-made software. 
The real-time positioning and intensity control of the heating laser, which is realized with an acousto-optic deflector (Brimrose, 2 DS-75-40-808), can be performed according to any protocol allowing for the investigation of a pluripotency of dynamic temperature fields \cite{braun2015single}. 
This technique is, thus, ideally suited for testing DeepCalib on experimental data obtained from a broad range of force fields.  

In this section, we use the thermophoretic trap to generate a restoring force field corresponding to a harmonic potential.
We record a 500-s trajectory ($5 \times 10^4$ samples, time step $10\,{\rm ms}$) and determine the ``true'' ground-truth $k$ by the variance method using the full recorded trajectory (black dashed lines in Figs.~\ref{fig2}b-d).
We then test the performance of DeepCalib on 400 (partially overlapping) segments of this trajectory (1000 samples each); the resulting estimations are presented by the orange histogram in Fig.~\ref{fig2}b (see also Example 1b of the DeepCalib software package \cite{DC}).
The estimations obtained by the variance and autocorrelation methods are presented by the blue and green histograms in Figs.~\ref{fig2}c and \ref{fig2}d, respectively, and show that these methods present a bias towards larger and smaller values of $k$, respectively.
Such biases can be explained by the short length of the trajectories. 
For the variance method, the trajectory is not long enough to explore the full potential well leading to an underestimation of the variance and, thus, an overestimation of $k$.
For the autocorrelation method, short trajectories exploring only the region near the equilibrium position lead to an overestimation of the correlation time in the trap and, thus, an underestimation of $k$.
Although DeepCalib is trained with simulated trajectories, it determines the trap stiffness from experimental trajectories more accurately than the standard methods:
DeepCalib estimations are both closer to the measured truth (lower bias) and less spread (higher precision).
Therefore, thanks to its data-driven training process, the RNN manages to combine the insight provided by the variance and autocorrelation methods, while largely avoiding their pitfalls.

\subsection{Double-well potential}

%%%%%%%%%%%%%%%%%%%%%%%%%%%%%
%% FIGURE 3
%%%%%%%%%%%%%%%%%%%%%%%%%%%%%
\begin{figure*} 
	\includegraphics[width= \textwidth]{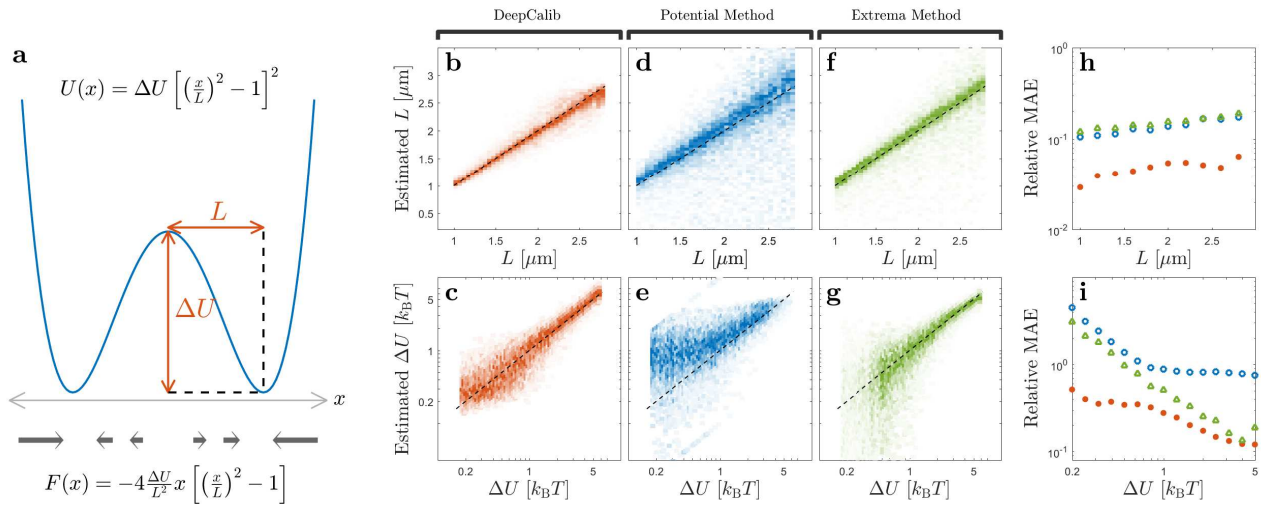}
	\caption{
		{\bf Calibration of bistable potentials (simulations).}
		{\bf a}, Potential energy profile (blue solid line) and force field (gray arrows in the bottom) of a double-well trap characterized by the equilibrium distance $Lž$ and the energy-barrier height $\Delta U$. 
		{\bf b}, Distribution of $L$ and, {\bf c}, $\Delta U$ estimated by DeepCalib from simulated trajectories; the black dashed line represents the ground truth. 
		{\bf d}--{\bf e}, Corresponding distributions estimated by the potential method and, {\bf f}--{\bf g}, by the extrema method (see details about these methods in the text).
		{\bf h}, Relative mean absolute error (MAE) of the estimations obtained using DeepCalib (orange dots), the potential method (blue circles), and the extrema method  (green triangles) as a function of $L$ and, {\bf i}, as a function of $\Delta U$: in both cases DeepCalib outperforms the other methods achieving lower MAE.
		These results are obtained from a test dataset of $10^4$ trajectories, each sampled 1000 times every $50\,{\rm ms}$. Both training and test trajectories are generated with $L$ values uniformly distributed in linear scale and $\Delta U$ values uniformly distributed in logarithmic scale.
		See also Example 2a of the DeepCalib software package \cite{DC}.
		}
		\label{fig3}
\end{figure*}
%%%%%%%%%%%%%%%%%%%%%%%%%%%%%

%%%%%%%%%%%%%%%%%%%%%%%%%%%%%
%% FIGURE 4
%%%%%%%%%%%%%%%%%%%%%%%%%%%%%
\begin{figure*}
	\includegraphics[width=.88 \textwidth]{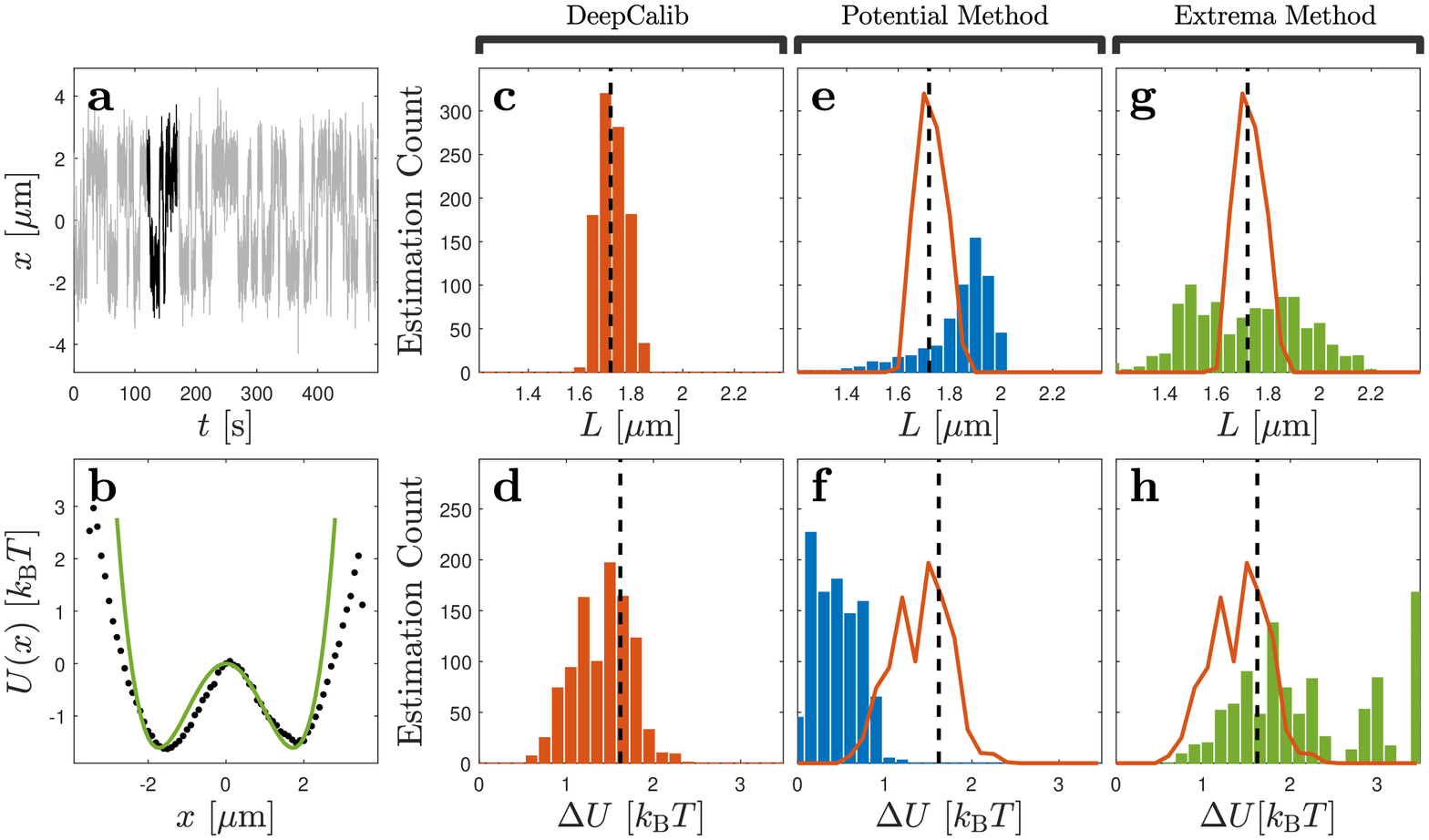}
	\caption{
		{\bf Calibration of bistable potentials (experiments).} 
		{\bf a}, Example of an experimental trajectory of a Brownian particle in a double-well potential (gray line). The highlighted section (black line) is an example of the trajectory portion used to estimate the potential parameters. 
		{\bf b}, The experimental potential energy landscape corresponding to the whole experimental trajectory in {\bf a} (black dots) and corresponding fitted potential using the extrema method (green line). Note the reality gap between the theory and the experiment \cite{cichos2020machine}. 
		{\bf c}-{\bf h}, Distributions of $L$ and $\Delta U$ estimated by DeepCalib (orange histograms in {\bf c} and {\bf d}, respectively, analyzed using the same RNN as in Fig.~\ref{fig3}), by the potential method (blue histograms in {\bf e} and {\bf f}, respectively), and by the extrema method (green histograms in {\bf g} and {\bf h}, respectively) for 900 50-s segments of a single 1500-s trajectory (each segment corresponds to 1000 samples taken every $50\,{\rm ms}$).
		The black dashed lines represent the estimations of $L$ ({\bf c}, {\bf e}, {\bf g}) and $\Delta U$ ({\bf d}, {\bf f}, {\bf h}) using the full length of the trajectory with the extrema method, which we take as ground truth.
		The orange lines in {\bf e} and {\bf g} ({\bf f} and {\bf h}) reproduce the histogram for DeepCalib from {\bf c} ({\bf d}) for ease of comparison: In all cases DeepCalib provides more accurate and precise estimations.
		See also Example 2b of the DeepCalib software package \cite{DC}.
		}
		\label{fig4}
\end{figure*}
%%%%%%%%%%%%%%%%%%%%%%%%%%%%%

Now that we have validated DeepCalib on the fundamental case of a harmonic trap, we move to the more complex case of a bistable potential.
Bistable traps represent a model system to study several physical and biological phenomena, such as Kramer'€™s transitions \cite{mccann1999thermally}, Landauer's principle \cite{jun2014high}, and folding energies of nucleic acids \cite{woodside2006direct}. 
The simplest analytic form for a double-well potential is given by a quartic polynomial (solid line in Fig.~\ref{fig3}a):
\begin{equation}
	U(x) 
	=  
	\Delta U 
	\left[
		\left(
			{x \over  L}
		\right)^2 
		- 1 
	\right]^2,
\end{equation}
where $x=\pm L$ are the local minima and $\Delta U$ is the barrier height. This gives rise  to a cubic force field (arrows in Fig.~\ref{fig3}a): 
\begin{equation}
	F(x) 
	= 
	-4
	{\Delta U \over L^2}  
	x 
	\left[
		\left(
			{x \over L}
		\right)^2 
		- 1 
	\right],
\end{equation} 
clearly showing that the force vanishes at the potential minima  $x=\pm L$ and at the local maximum $x=0$.
The parameters that characterize this double-well potential are the equilibrium distance $L$ and the energy barrier height $\Delta U$. 

The RNN employed by DeepCalib is similar to that for the harmonic trap case, but having two outputs to estimate both $L$ and $\Delta U$.
We train this RNN on about $ 10^7 $ trajectories that are simulated with $\Delta U$ ranging from 0.1 to 10 $ k_{\rm B}T $ (uniformly distributed in logarithmic scale) and $L$ ranging from 1 $\rm \upmu m$ to 3 $\rm \upmu m$ (uniformly distributed in linear scale).
Finally, we test its performance on 10000 simulated trajectories with 1000 samples (time step $50\,{\rm ms}$).
DeepCalib provides accurate estimations for both $L$ (orange distribution, Fig.~\ref{fig3}b) and $\Delta U$ (orange distribution, Fig.~\ref{fig3}c) for a wide range of parameters (the ground truth is plotted by the black dashed lines). 
More details can be found in Example 2a of the DeepCalib software package \cite{DC}.

We compare the performance of DeepCalib (Figs.~\ref{fig3}b--c) to standard methods (Figs.~\ref{fig3}d--g). 
The standard methods to calibrate a double-well potential use the relation between equilibrium probability distribution and the potential energy \cite{jones2015optical}, which is given by
\begin{equation}	
	\rho(x)
	= 
	\frac{
		e^{
			-U(x)
			/
			k_{\rm B}T
		}
	}{
		N
	},
\end{equation}
where the normalization factor $N = \int_{-\infty}^{\infty} e^{- U(x)/k_{\rm B}T} {\rm d}x$ is the partition function.
Here, we use two concrete approaches.
First, we perform a quartic fit to $\ln \rho(x)$ to determine the optimal values of $L$ and $\Delta U$ (``potential method''\cite{berut2012experimental}, Figs.~\ref{fig3}d--e). However, we observe that, for short trajectories, the potential method estimates $ \Delta U$ with a strong bias. 
Thus, we employ a second method that is more accurate for shorter trajectories: As $\rho(x)$ displays two local maxima at $\pm L$ (potential minima) and a local minimum at the origin (potential barrier), we obtain $L$ as the distance between the maximum of $\rho(x)$ and the origin, and $\Delta U$ as the ratio of the maximum probability and the probability at the origin (``extrema method'' \cite{mccann1999thermally}, Figs.~\ref{fig3}f--g). Although providing much better estimations than the potential method, also the extrema method achieves a significantly worse performance than DeepCalib because of the limited length of the trajectories.
This is confirmed by the inspection of the relative MAE (Figs.~\ref{fig3}h--i): The relative MAE of DeepCalib (orange dots) is much lower than that of the potential method (blue circles) and of the extrema method (green triangles) over the whole range of both $L$ (Fig.~\ref{fig3}h) and $\Delta U$ (Fig.~\ref{fig3}i).

Finally, we test the performance of DeepCalib on experimental trajectories while using the same RNN employed for the analysis of the simulated data.
The experimental data are acquired using the same thermophoretic setup employed for the harmonic trap (Fig.~\ref{fig2}a), but imposing the force field of a double-well trap. 
We record a 1500-s trajectory (150000 samples, time step $10\,{\rm ms}$). A part of the experimental trajectory is shown in Fig.~\ref{fig4}a. Interestingly, the experimental potential is not exactly a quartic potential (a typical example of the ``reality gap'' (Fig.~\ref{fig4}b) between experiments and simulations \cite{cichos2020machine}). 
The experimental potential obtained with the full extent of the trajectory is shown in Fig.~\ref{fig4}b. We determine the ``true'' ground-truth values for $L$ and $\Delta U$ using the extrema method (green line Fig.~\ref{fig4}b and black dashed lines, Fig.~\ref{fig4}c--h). 
This reality gap makes it particularly interesting to assess how the various methods perform, because DeepCalib is trained on the idealized quartic potential, and the potential method assumes a quartic potential in its analysis. 
We test the performance of DeepCalib on 900 (partially overlapping) segments of this trajectory (1000 samples each with time step $50\,{\rm ms}$, highlighted black line in \ref{fig4}a) obtaining the estimations of $L$ and $\Delta U$ represented by the orange histograms in Figs.~\ref{fig4}c and \ref{fig4}d, respectively (see also Example 2b of the DeepCalib software package \cite{DC}). The corresponding estimations for the potential method are provided by the blue histograms in Figs.~\ref{fig4}e--f, and those for the extrema method by the green histograms in Figs.~\ref{fig4}g--h.
Also in this case, DeepCalib is more accurate and less biased than the standard methods.
In particular, we highlight the fact that DeepCalib provides accurate estimations even though the experimental potential differs from the idealized double-well potential employed in the simulations used in its training.
This demonstrates that the neural-network approach put forward by DeepCalib can efficiently bridge the reality gap between idealized simulations and actual experiments.

\subsection{Rotational force field}

%%%%%%%%%%%%%%%%%%%%%%%%%%%%%
%% FIGURE 5
%%%%%%%%%%%%%%%%%%%%%%%%%%%%%
\begin{figure*} 
	\includegraphics[width= \textwidth]{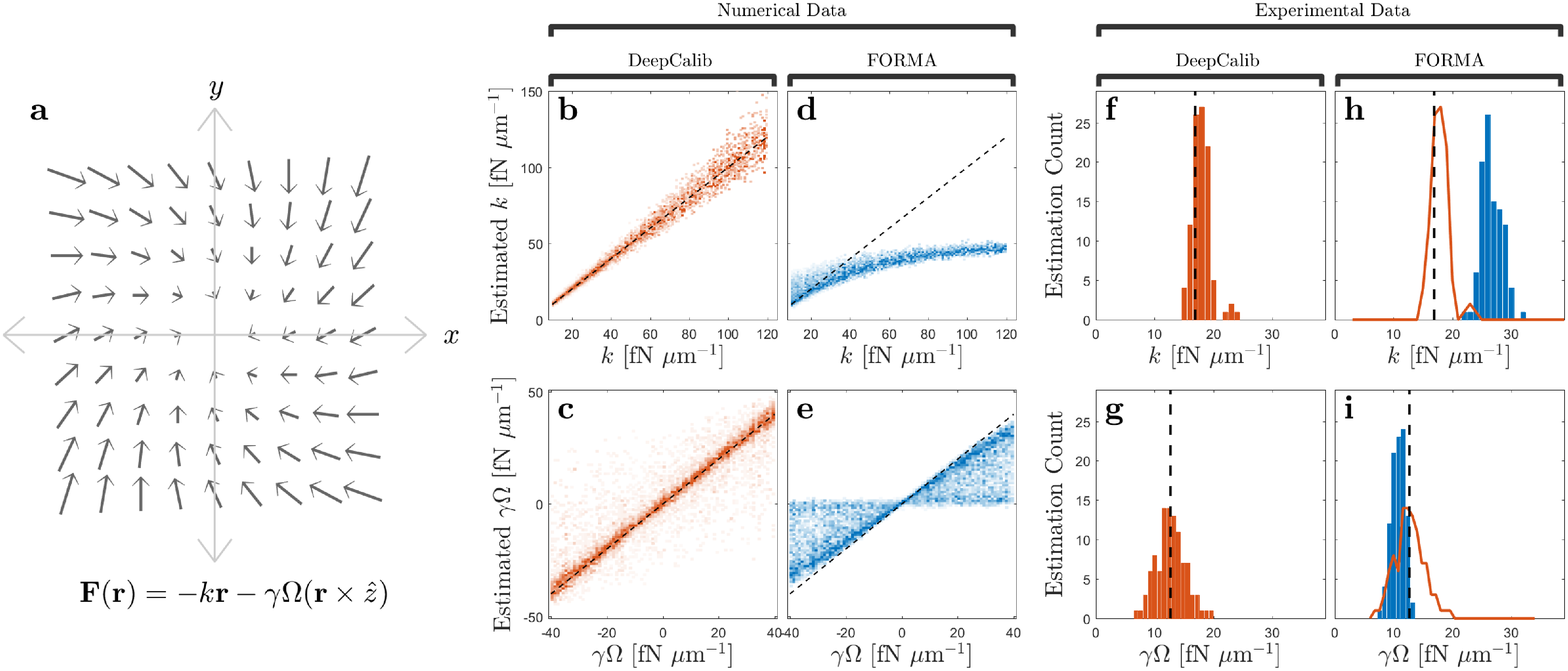}
	\caption{
		{\bf Calibration of a non-conservative force field.}
		{\bf a}, Non-conservative force field consisting of a harmonic potential characterized by the stiffness $k$ and a rotational force field characterized by the rotational parameter $ \Omega$. 
		{\bf b}--{\bf e}, Distributions of $k$ and $\gamma \Omega$
		estimated by DeepCalib ({\bf b} and {\bf c}, respectively) and by FORMA  ({\bf d} and {\bf e}, respectively) from simulated trajectories; the black dashed lines represent the ground truth.
		Both training and test trajectories are generated with $k$ values uniformly distributed in logarithmic scale and $\Omega$ values uniformly distributed in linear scale.
		These results are obtained from a test dataset of $10^4$ trajectories, each sampled 1000 times every $50\,{\rm ms}$.
		{\bf f}--{\bf i}, Distributions of $k$ and $\Omega$ estimated by DeepCalib (orange histograms in {\bf f} and {\bf g}, respectively, analyzed using the same RNN as in {\bf b}--{\bf e}) and by FORMA  (blue histograms in {\bf h} and {\bf i}, respectively) for 100 50-s segments of a single 1000-s trajectory (each segment corresponds to 1000 samples taken every $50\,{\rm ms}$). The black dashed lines represents the FORMA-based estimations of $k$ ({\bf f}, {\bf h}) and $\Omega$ ({\bf g}, {\bf i}) using the full length of the trajectory sampled every $10\,{\rm ms}$, which we take as ground truth.
		See also Examples 3a and 3b of the DeepCalib software package \cite{DC}.
	}
	\label{fig5}
\end{figure*}
%%%%%%%%%%%%%%%%%%%%%%%%%%%%%

We now test DeepCalib in a non-equilibrium scenario created by a non-conservative rotational force field. 
Non-conservative force fields are widely used  to investigate the non-equilibrium dynamics and thermodynamics of microscopic systems   
\cite{volpe2006torque, volpe2007brownian, blickle2007einstein, gomez2009experimental}. We consider the rotational force field described by the following equation: 
\begin{equation}
	{\bf F}({\bf r}) 
	= 
	-k{\bf r} 
	-\gamma \Omega
	({\bf r} \times \hat{\bf z}),
\end{equation}
where $\bf{r}$ is the two-dimensional position in the $xy$-plane of the Brownian particle, which is subject to a restoring force with stiffness $k$ and a torque with rotational frequency $\Omega$.
An example of a rotational force field is shown in Fig.~\ref{fig5}a.
This non-equilibrium system relaxes to a steady state, but its distribution is determined only by the restoring force and is independent of $\Omega$ ($\rho(x,y) \propto e^{-k(x^2+y^2)/T}$ \cite{volpe2006torque}).
Thus, differently from the previous examples, even in principle, it is impossible to use the steady-state probability distribution to calibrate this force field, regardless of the amount of available data.
The available methods \cite{volpe2006torque, volpe2007brownian, garcia2018high, frishman2020learning} rely essentially on local drifts and, therefore, require high-frequency measurements (i.e., the measurement time step must be at least one order of magnitude smaller than the characteristic times associated to the motion of the Brownian particle in the force field, which in this case are $\tau_{\rm c} = \gamma / k$ and $\tau_{\rm r} = \Omega^{-1}$ \cite{volpe2006torque, volpe2007brownian}).

Also for this example, we train DeepCalib on simulated trajectories with 1000 samples acquired with a time step of $50\,{\rm ms}$, but in this case we use two-dimensional trajectories. We train this RNN on about $ 10^7 $ trajectories that are simulated with $k$ ranging from $6$ to $150\,{\rm fN \upmu m}^{-1}$ (uniformly distributed in logarithmic scale) and $\gamma \Omega$ ranging from $-42$ to $42\,{\rm fN \upmu m}^{-1}$ (uniformly distributed in linear scale).

DeepCalib manages to estimate with good accuracy both $k$ and $\Omega$, as can be seen by comparing the orange distributions and the ground-truth values provided by the black dashed lines in Figs.~\ref{fig5}b and \ref{fig5}c, respectively (see also Example 3a of the DeepCalib software package \cite{DC}).

Since the time step is comparable to the characteristic time of the system, we expect the standard methods to fail \cite{garcia2018high, frishman2020learning}.
In fact, when we apply FORMA \cite{garcia2018high} to calibrate this force field, we obtain much poorer estimations (blue distributions in Figs.~\ref{fig5}d and \ref{fig5}e).
FORMA performs reasonably well for low $k$ (longer characteristic times), but fails for higher values of $k$ (shorter characteristic times), while it performs poorly over the whole range of $\Omega$.

Finally, we test the performance of DeepCalib for an experimental rotational force field, generated using the thermophoteric setup (Fig.~\ref{fig2}a). We make the test on 500 (partially overlapping) segments of the experimental trajectory (1000 seconds long), each with 1000 samples with the time step of 50 ms. 
The estimation of the force-field parameters is challenging because the 50-ms measurement time step is comparable to the force-field characteristic times ($\tau_{\rm c} = 145\,{\rm ms}$, $\tau_{\rm r} = 193\,{\rm ms}$).
We determine the ``true'' ground-truth values of $k$ and $\Omega$ (black dashed lines in Figs.~\ref{fig5}f--i) with the FORMA-based estimations using the full length of the trajectory sampled more often (i.e., every $10\,{\rm ms}$ instead of every $50\,{\rm ms}$), so that the sampling time is much shorter that $\tau_{\rm c}$ and $\tau_{\rm r}$.
Once again, the estimations of $k$ by DeepCalib (orange distribution, Fig.~\ref{fig5}f) are more accurate than those by FORMA (blue distribution, Fig.~\ref{fig5}h), which clearly deviate from the measured ground truth (black dashed lines). 
Likewise, the estimations of $\Omega$ by DeepCalib (orange distribution, Fig.~\ref{fig5}g) are also closer to the measured truth (black dashed lines) than those by FORMA (blue distribution, Fig.~\ref{fig5}j).
For further details, see also Example 3b of the DeepCalib software package \cite{DC}.

\subsection{Dynamical nonequilibrium trap}

%%%%%%%%%%%%%%%%%%%%%%%%%%%%%
%% FIGURE 6
%%%%%%%%%%%%%%%%%%%%%%%%%%%%%
\begin{figure*} [ht!]
	\includegraphics[width= \textwidth]{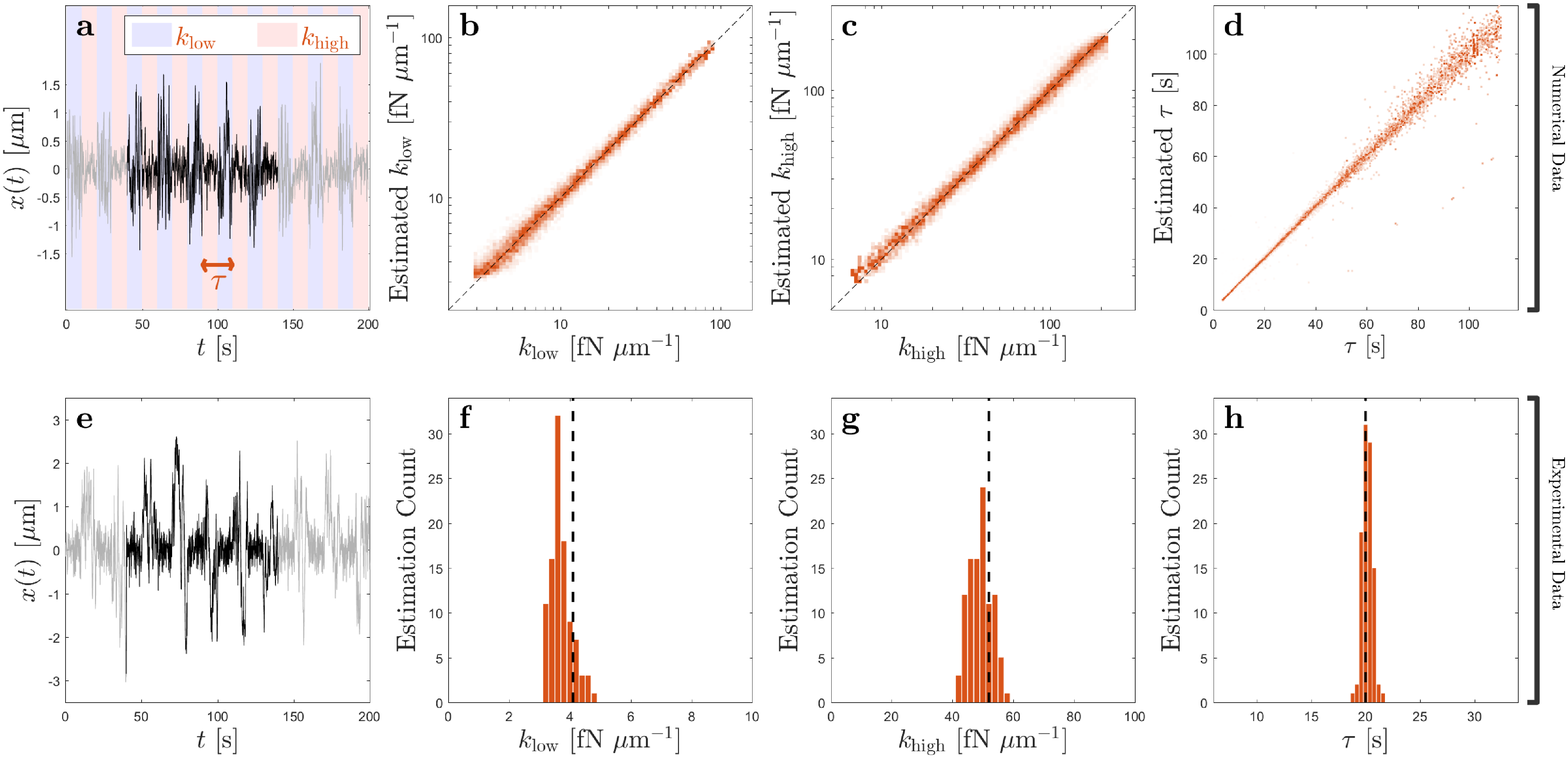}
	\caption{
		{\bf Calibration of a time-varying force field.}
		{\bf a}, Trajectory of a Brownian particle in a harmonic trap whose stiffness $k$ switches over time between a lower stiffness value $k_{\rm low}$ and a higher stiffness value $k_{\rm high}$ with a period of $\tau$.
		{\bf b}--{\bf d}, Distribution of $k_{\rm low}$, $k_{\rm high}$ and $\tau$ estimated by DeepCalib as a function of their ground truth (black dashed lines) for $2\times10^4$ simulated trajectories with 1000 samples taken every $100\,{\rm ms}$.
		Both training and test trajectories are generated with $k_{\rm low}$, $k_{\rm high}$ and $\tau$ values uniformly distributed in logarithmic scale. 
		{\bf e}, Experimental trajectory of a Brownian particle in a harmonic trap whose stiffness $k$ switches over time between $k_{\rm low} = 4.1$ ${\rm fN} \mathrm{\upmu m}^{-1}$ and $k_{\rm high} = 52$ ${\rm fN} \mathrm{\upmu m}^{-1}$ with a period of $\tau = 20$ ${\rm ms}$.
		{\bf f}-{\bf h}, Histograms of $k_{\rm low}$, $k_{\rm high}$ and $\tau$ estimated by DeepCalib for 100 100-s segments of a single experimental trajectory (each segment corresponds to 1000 samples taken every $100\,{\rm ms}$) analyzed using the same RNN as in {\bf c}--{\bf e}. 
		The black dashed lines in {\bf f}--{\bf h} represent the ground truth values of $k_{\rm low}$, $k_{\rm high}$ and $\tau$ ($k_{\rm low}$ and $k_{\rm high}$ are measured from recorded trajectories kept at constant stiffness, and $\tau$ is the set period in the switching of the protocol): DeepCalib achieves estimations of these parameters with low variance even using very short trajectory segments.	
		See also Examples 4a and 4b of the DeepCalib software package \cite{DC}.
	}
	\label{fig6}
\end{figure*}
%%%%%%%%%%%%%%%%%%%%%%%%%%%%%

To further demonstrate the potentiality of DeepCalib, we set to calibrate an even more challenging dynamical nonequilibrium system. 
We consider a Brownian particle subject to an alternating trapping potential that is switching between a low stiffness $k_{\rm low}$ and a high stiffness $k_{\rm high}$ with a period $\tau$.
Fig.~\ref{fig6}a shows an example trajectory together with the corresponding stiffness protocol. 
There is no simple standard method for calibrating such a system, as one would have to combine techniques to detect the switching points (see, e.g., \cite{montiel2006quantitative}) with techniques to estimate  stiffnesses (such as the variance and autocorrelation method that we discussed for the harmonic trap) on shorter segments of the trajectory. However, it is quite difficult to estimate these parameters for most cases as the exact switching point gets very difficult to determine when the stiffness values are close (Fig.~\ref{fig6}a features an example with a large difference between $k_{\rm low}$ and $k_{\rm high}$). In addition, as the system is continuously kept in a nonequilibrium state, the variance and the autocorrelation methods cannot be used.  

DeepCalib can be straightforwardly applied also to this case.
We train DeepCalib on simulated trajectories with 1000 samples acquired with a time step of $100\,{\rm ms}$. We train this RNN on about $ 10^7 $ trajectories that are simulated with $k_{\rm low}$ and $k_{\rm high}$ ranging from 2 to 280 fN $\mathrm{\rm \upmu m}^{-1}$ (uniformly distributed in logarithmic scale, with a condition that $k_{\rm high} > 2k_{\rm low}$) and $\tau$ ranging from $3$ to $110\,{\rm s}$ (uniformly distributed in logarithmic scale).
We then test the trained RNN on $2 \times 10^4$ simulated trajectories, demonstrating that it is able to simultaneously and accurately estimate $k_{\rm low}$ (Fig.~\ref{fig6}b), $k_{\rm high}$ (Fig.~\ref{fig6}c) and $\tau$ (Fig.~\ref{fig6}d) (see also Example 4a of the DeepCalib software package \cite{DC}). 

Experimentally, we realize this protocol using a thermophoretic harmonic trap that alternates between two stiffnesses. We record the experimental trajectory of $10^5$ data samples with 10 ms time step, a part of this trajectory is shown in Fig.~\ref{fig6}e. We then make the test on 100 (partially overlapping) segments of the experimental trajectory each with 1000 samples with the time step of $100 \,{\rm ms}$ (black line, Fig.~\ref{fig6}e). 
The measured ground truth (black dashed lines in Figs.~\ref{fig6}f--h) for the stiffnesses of the experimental data is obtained from trajectories recorded at constant stiffnesses $k_{\rm low}$ and $k_{\rm high}$, while we know exactly the ground truth for $\tau$ because we control the period of the experimental switching of the protocol.
Using the same RNN trained for Figs.~\ref{fig6}b--d, DeepCalib successfully estimates the parameters of the system  $k_{\rm low}$ (Fig.~\ref{fig6}f), $k_{\rm high}$ (Fig.~\ref{fig6}g) and $\tau$ (Fig.~\ref{fig6}h) from the experimental data (see also Example 4b of the DeepCalib software package \cite{DC}). 

This latter example demonstrates that DeepCalib can be directly applied to rather generic settings 
beyond simple equilibrium or steady-state dynamics, for which standard techniques are not available and one would have to develop system-specific analysis methods.

\section{DeepCalib Software Package}

We provide DeepCalib on GitHub as a Python open-source freeware software package, which can be readily personalized and optimized for the needs of specific users and applications \cite{DC}. 
The user can easily adapt DeepCalib to the analysis of any force field by altering the stochastic differential equations describing the motion of the Brownian particle used for the simulation of the training datasets.
This gives users the ability to train their own RNN in order to calibrate their specific force field with no prior machine learning knowledge. 
The trained RNN can also be saved to be used on other software platforms (e.g., MATLAB and LabVIEW). 
This opens the possibility to straightforwardly analyze any force field, even when no standard calibration techniques are available, greatly enhancing the range of microscopic systems that can be analyzed and studied.
 
\section{Conclusion}

We have introduced DeepCalib, a data-driven neural-network approach for the calibration of microscopic force fields acting on a Brownian particle, and reported its performance.
By benchmarking it on simple tasks, for which standard techniques are available, we have shown that it outperforms standard methods in challenging conditions involving short and/or low frequency measurements. 
Then, we have demonstrated that it can be straightforwardly applied to non-equilibrium, unsteady force fields, for which no simple standard technique exists.
We have also demonstrated that DeepCalib, while trained on simulated data, is able to generalize and successfully calibrate force fields from experimental data. 
Remarkably, even when the model of the force field used for the training was not perfectly matching the experimental one, as in the case of the double trap, DeepCalib managed to extract the key features such as the location of the traps and the barrier height better than the standard methods. 
This demonstrates its capability of bridging the reality gap between the idealized simulation used for training and the experimental reality.

DeepCalib is thus a flexible method that can be used on a wide variety of calibration tasks. 
This can be clearly appreciated by considering that there is no standard technique that we could have used to address all the examples we have considered. 
Indeed, even for the scenarios that admit standard methods, we had to employ different methods for each case, whereas DeepCalib just needed different training sets and the minor modification of adjusting the number of outputs to match the number of the desired calibration parameters.
Therefore, DeepCalib is ideal to calibrate complex and non-standard force fields from short trajectories, for which advanced specific method would have to be developed on a case-by-case basis. 
Potential areas of application include  the real time calibration of bistable potentials used for information theory \cite{jun2014high}, improvement of the analysis of microscopic heat engines \cite{martinez2017colloidal}, and prediction of the free energies of biomolecules \cite{hummer2010free}.

\begin{acknowledgments}
The authors thank Harshith Bachimanchi and Martin Selin for critically revising the manuscript and the software.
\end{acknowledgments}

\bibliography{biblio}

\begin{thebibliography}{64}%
\makeatletter
\providecommand \@ifxundefined [1]{%
 \@ifx{#1\undefined}
}%
\providecommand \@ifnum [1]{%
 \ifnum #1\expandafter \@firstoftwo
 \else \expandafter \@secondoftwo
 \fi
}%
\providecommand \@ifx [1]{%
 \ifx #1\expandafter \@firstoftwo
 \else \expandafter \@secondoftwo
 \fi
}%
\providecommand \natexlab [1]{#1}%
\providecommand \enquote  [1]{``#1''}%
\providecommand \bibnamefont  [1]{#1}%
\providecommand \bibfnamefont [1]{#1}%
\providecommand \citenamefont [1]{#1}%
\providecommand \href@noop [0]{\@secondoftwo}%
\providecommand \href [0]{\begingroup \@sanitize@url \@href}%
\providecommand \@href[1]{\@@startlink{#1}\@@href}%
\providecommand \@@href[1]{\endgroup#1\@@endlink}%
\providecommand \@sanitize@url [0]{\catcode `\\12\catcode `\$12\catcode
  `\&12\catcode `\#12\catcode `\^12\catcode `\_12\catcode `\%12\relax}%
\providecommand \@@startlink[1]{}%
\providecommand \@@endlink[0]{}%
\providecommand \url  [0]{\begingroup\@sanitize@url \@url }%
\providecommand \@url [1]{\endgroup\@href {#1}{\urlprefix }}%
\providecommand \urlprefix  [0]{URL }%
\providecommand \Eprint [0]{\href }%
\providecommand \doibase [0]{http://dx.doi.org/}%
\providecommand \selectlanguage [0]{\@gobble}%
\providecommand \bibinfo  [0]{\@secondoftwo}%
\providecommand \bibfield  [0]{\@secondoftwo}%
\providecommand \translation [1]{[#1]}%
\providecommand \BibitemOpen [0]{}%
\providecommand \bibitemStop [0]{}%
\providecommand \bibitemNoStop [0]{.\EOS\space}%
\providecommand \EOS [0]{\spacefactor3000\relax}%
\providecommand \BibitemShut  [1]{\csname bibitem#1\endcsname}%
\let\auto@bib@innerbib\@empty
%</preamble>
\bibitem [{\citenamefont {Jones}\ \emph {et~al.}(2015)\citenamefont {Jones},
  \citenamefont {Marag\`o},\ and\ \citenamefont {Volpe}}]{jones2015optical}%
  \BibitemOpen
  \bibfield  {author} {\bibinfo {author} {\bibfnamefont {P.~H.}\ \bibnamefont
  {Jones}}, \bibinfo {author} {\bibfnamefont {O.~M.}\ \bibnamefont {Marag\`o}},
  \ and\ \bibinfo {author} {\bibfnamefont {G.}~\bibnamefont {Volpe}},\
  }\href@noop {} {\emph {\bibinfo {title} {Optical tweezers: Principles and
  applications}}}\ (\bibinfo  {publisher} {Cambridge University},\ \bibinfo
  {year} {2015})\BibitemShut {NoStop}%
\bibitem [{\citenamefont {Wu}(2011)}]{wu2011optoelectronic}%
  \BibitemOpen
  \bibfield  {author} {\bibinfo {author} {\bibfnamefont {M.~C.}\ \bibnamefont
  {Wu}},\ }\bibfield  {title} {\enquote {\bibinfo {title} {Optoelectronic
  tweezers},}\ }\href@noop {} {\bibfield  {journal} {\bibinfo  {journal} {Nat.
  Photon.}\ }\textbf {\bibinfo {volume} {5}},\ \bibinfo {pages} {322} (\bibinfo
  {year} {2011})}\BibitemShut {NoStop}%
\bibitem [{\citenamefont {Braun}\ and\ \citenamefont
  {Cichos}(2013)}]{braun2013optically}%
  \BibitemOpen
  \bibfield  {author} {\bibinfo {author} {\bibfnamefont {M.}~\bibnamefont
  {Braun}}\ and\ \bibinfo {author} {\bibfnamefont {F.}~\bibnamefont {Cichos}},\
  }\bibfield  {title} {\enquote {\bibinfo {title} {Optically controlled
  thermophoretic trapping of single nano-objects},}\ }\href@noop {} {\bibfield
  {journal} {\bibinfo  {journal} {ACS Nano}\ }\textbf {\bibinfo {volume} {7}},\
  \bibinfo {pages} {11200--11208} (\bibinfo {year} {2013})}\BibitemShut
  {NoStop}%
\bibitem [{\citenamefont {Gieseler}\ \emph {et~al.}(2020)\citenamefont
  {Gieseler}, \citenamefont {Gomez-Solano}, \citenamefont {Magazz{\`u}},
  \citenamefont {Castillo}, \citenamefont {Garc{\'\i}a}, \citenamefont
  {Gironella-Torrent}, \citenamefont {Viader-Godoy}, \citenamefont {Ritort},
  \citenamefont {Pesce}, \citenamefont {Arzola} \emph
  {et~al.}}]{gieseler2020optical}%
  \BibitemOpen
  \bibfield  {author} {\bibinfo {author} {\bibfnamefont {J.}~\bibnamefont
  {Gieseler}}, \bibinfo {author} {\bibfnamefont {J.~R.}\ \bibnamefont
  {Gomez-Solano}}, \bibinfo {author} {\bibfnamefont {A.}~\bibnamefont
  {Magazz{\`u}}}, \bibinfo {author} {\bibfnamefont {I.~P.}\ \bibnamefont
  {Castillo}}, \bibinfo {author} {\bibfnamefont {L.~P.}\ \bibnamefont
  {Garc{\'\i}a}}, \bibinfo {author} {\bibfnamefont {M.}~\bibnamefont
  {Gironella-Torrent}}, \bibinfo {author} {\bibfnamefont {X.}~\bibnamefont
  {Viader-Godoy}}, \bibinfo {author} {\bibfnamefont {F.}~\bibnamefont
  {Ritort}}, \bibinfo {author} {\bibfnamefont {G.}~\bibnamefont {Pesce}},
  \bibinfo {author} {\bibfnamefont {A.~V.}\ \bibnamefont {Arzola}},  \emph
  {et~al.},\ }\bibfield  {title} {\enquote {\bibinfo {title} {Optical tweezers:
  A comprehensive tutorial from calibration to applications},}\ }\href@noop {}
  {\bibfield  {journal} {\bibinfo  {journal} {arXiv preprint arXiv:2004.05246}\
  } (\bibinfo {year} {2020})}\BibitemShut {NoStop}%
\bibitem [{\citenamefont {Mills}\ \emph {et~al.}(2004)\citenamefont {Mills},
  \citenamefont {Qie}, \citenamefont {Dao}, \citenamefont {Lim}, \citenamefont
  {Suresh} \emph {et~al.}}]{mills2004nonlinear}%
  \BibitemOpen
  \bibfield  {author} {\bibinfo {author} {\bibfnamefont {J.~P.}\ \bibnamefont
  {Mills}}, \bibinfo {author} {\bibfnamefont {L.}~\bibnamefont {Qie}}, \bibinfo
  {author} {\bibfnamefont {M.}~\bibnamefont {Dao}}, \bibinfo {author}
  {\bibfnamefont {C.~T.}\ \bibnamefont {Lim}}, \bibinfo {author} {\bibfnamefont
  {S.}~\bibnamefont {Suresh}},  \emph {et~al.},\ }\bibfield  {title} {\enquote
  {\bibinfo {title} {Nonlinear elastic and viscoelastic deformation of the
  human red blood cell with optical tweezers},}\ }\href@noop {} {\bibfield
  {journal} {\bibinfo  {journal} {Mech. Chem. Biosys.}\ }\textbf {\bibinfo
  {volume} {1}},\ \bibinfo {pages} {169--180} (\bibinfo {year}
  {2004})}\BibitemShut {NoStop}%
\bibitem [{\citenamefont {Sleep}\ \emph {et~al.}(1999)\citenamefont {Sleep},
  \citenamefont {Wilson}, \citenamefont {Simmons},\ and\ \citenamefont
  {Gratzer}}]{sleep1999elasticity}%
  \BibitemOpen
  \bibfield  {author} {\bibinfo {author} {\bibfnamefont {J.}~\bibnamefont
  {Sleep}}, \bibinfo {author} {\bibfnamefont {D.}~\bibnamefont {Wilson}},
  \bibinfo {author} {\bibfnamefont {R.}~\bibnamefont {Simmons}}, \ and\
  \bibinfo {author} {\bibfnamefont {W.}~\bibnamefont {Gratzer}},\ }\bibfield
  {title} {\enquote {\bibinfo {title} {Elasticity of the red cell membrane and
  its relation to hemolytic disorders: an optical tweezers study},}\
  }\href@noop {} {\bibfield  {journal} {\bibinfo  {journal} {Biophys. J.}\
  }\textbf {\bibinfo {volume} {77}},\ \bibinfo {pages} {3085--3095} (\bibinfo
  {year} {1999})}\BibitemShut {NoStop}%
\bibitem [{\citenamefont {Su}\ \emph {et~al.}(2003)\citenamefont {Su},
  \citenamefont {Wei}, \citenamefont {Zhang}, \citenamefont {Mock},
  \citenamefont {Smith},\ and\ \citenamefont {Schultz}}]{su2003interparticle}%
  \BibitemOpen
  \bibfield  {author} {\bibinfo {author} {\bibfnamefont {K.~H.}\ \bibnamefont
  {Su}}, \bibinfo {author} {\bibfnamefont {Q.~H.}\ \bibnamefont {Wei}},
  \bibinfo {author} {\bibfnamefont {X.}~\bibnamefont {Zhang}}, \bibinfo
  {author} {\bibfnamefont {J.~J.}\ \bibnamefont {Mock}}, \bibinfo {author}
  {\bibfnamefont {D.~R.}\ \bibnamefont {Smith}}, \ and\ \bibinfo {author}
  {\bibfnamefont {S.}~\bibnamefont {Schultz}},\ }\bibfield  {title} {\enquote
  {\bibinfo {title} {Interparticle coupling effects on plasmon resonances of
  nanogold particles},}\ }\href@noop {} {\bibfield  {journal} {\bibinfo
  {journal} {Nano Lett.}\ }\textbf {\bibinfo {volume} {3}},\ \bibinfo {pages}
  {1087--1090} (\bibinfo {year} {2003})}\BibitemShut {NoStop}%
\bibitem [{\citenamefont {Yada}\ \emph {et~al.}(2004)\citenamefont {Yada},
  \citenamefont {Yamamoto},\ and\ \citenamefont {Yokoyama}}]{yada2004direct}%
  \BibitemOpen
  \bibfield  {author} {\bibinfo {author} {\bibfnamefont {M.}~\bibnamefont
  {Yada}}, \bibinfo {author} {\bibfnamefont {J.}~\bibnamefont {Yamamoto}}, \
  and\ \bibinfo {author} {\bibfnamefont {H.}~\bibnamefont {Yokoyama}},\
  }\bibfield  {title} {\enquote {\bibinfo {title} {Direct observation of
  anisotropic interparticle forces in nematic colloids with optical
  tweezers},}\ }\href@noop {} {\bibfield  {journal} {\bibinfo  {journal} {Phys.
  Rev. Lett.}\ }\textbf {\bibinfo {volume} {92}},\ \bibinfo {pages} {185501}
  (\bibinfo {year} {2004})}\BibitemShut {NoStop}%
\bibitem [{\citenamefont {Paladugu}\ \emph {et~al.}(2016)\citenamefont
  {Paladugu}, \citenamefont {Callegari}, \citenamefont {Tuna}, \citenamefont
  {Barth}, \citenamefont {Dietrich}, \citenamefont {Gambassi},\ and\
  \citenamefont {Volpe}}]{paladugu2016nonadditivity}%
  \BibitemOpen
  \bibfield  {author} {\bibinfo {author} {\bibfnamefont {S.}~\bibnamefont
  {Paladugu}}, \bibinfo {author} {\bibfnamefont {A.}~\bibnamefont {Callegari}},
  \bibinfo {author} {\bibfnamefont {Y.}~\bibnamefont {Tuna}}, \bibinfo {author}
  {\bibfnamefont {L.}~\bibnamefont {Barth}}, \bibinfo {author} {\bibfnamefont
  {S.}~\bibnamefont {Dietrich}}, \bibinfo {author} {\bibfnamefont
  {A.}~\bibnamefont {Gambassi}}, \ and\ \bibinfo {author} {\bibfnamefont
  {G.}~\bibnamefont {Volpe}},\ }\bibfield  {title} {\enquote {\bibinfo {title}
  {Nonadditivity of critical {C}asimir forces},}\ }\href@noop {} {\bibfield
  {journal} {\bibinfo  {journal} {Nat. Commun.}\ }\textbf {\bibinfo {volume}
  {7}},\ \bibinfo {pages} {11403} (\bibinfo {year} {2016})}\BibitemShut
  {NoStop}%
\bibitem [{\citenamefont {Liphardt}\ \emph {et~al.}(2002)\citenamefont
  {Liphardt}, \citenamefont {Dumont}, \citenamefont {Smith}, \citenamefont
  {Jr.},\ and\ \citenamefont {Bustamante}}]{liphardt2002equilibrium}%
  \BibitemOpen
  \bibfield  {author} {\bibinfo {author} {\bibfnamefont {J.}~\bibnamefont
  {Liphardt}}, \bibinfo {author} {\bibfnamefont {S.}~\bibnamefont {Dumont}},
  \bibinfo {author} {\bibfnamefont {S.~B.}\ \bibnamefont {Smith}}, \bibinfo
  {author} {\bibfnamefont {I.~Tinoco}\ \bibnamefont {Jr.}}, \ and\ \bibinfo
  {author} {\bibfnamefont {C.}~\bibnamefont {Bustamante}},\ }\bibfield  {title}
  {\enquote {\bibinfo {title} {Equilibrium information from nonequilibrium
  measurements in an experimental test of {Jarzynski}'s equality},}\
  }\href@noop {} {\bibfield  {journal} {\bibinfo  {journal} {Science}\ }\textbf
  {\bibinfo {volume} {296}},\ \bibinfo {pages} {1832--1835} (\bibinfo {year}
  {2002})}\BibitemShut {NoStop}%
\bibitem [{\citenamefont {Collin}\ \emph {et~al.}(2005)\citenamefont {Collin},
  \citenamefont {Ritort}, \citenamefont {Jarzynski}, \citenamefont {Smith},
  \citenamefont {Tinoco},\ and\ \citenamefont
  {Bustamante}}]{collin2005verification}%
  \BibitemOpen
  \bibfield  {author} {\bibinfo {author} {\bibfnamefont {D.}~\bibnamefont
  {Collin}}, \bibinfo {author} {\bibfnamefont {F.}~\bibnamefont {Ritort}},
  \bibinfo {author} {\bibfnamefont {C.}~\bibnamefont {Jarzynski}}, \bibinfo
  {author} {\bibfnamefont {S.~B.}\ \bibnamefont {Smith}}, \bibinfo {author}
  {\bibfnamefont {I.}~\bibnamefont {Tinoco}}, \ and\ \bibinfo {author}
  {\bibfnamefont {C.}~\bibnamefont {Bustamante}},\ }\bibfield  {title}
  {\enquote {\bibinfo {title} {Verification of the {C}rooks fluctuation theorem
  and recovery of {RNA} folding free energies},}\ }\href@noop {} {\bibfield
  {journal} {\bibinfo  {journal} {Nature}\ }\textbf {\bibinfo {volume} {437}},\
  \bibinfo {pages} {231--234} (\bibinfo {year} {2005})}\BibitemShut {NoStop}%
\bibitem [{\citenamefont {Jun}\ \emph {et~al.}(2014)\citenamefont {Jun},
  \citenamefont {Gavrilov},\ and\ \citenamefont {Bechhoefer}}]{jun2014high}%
  \BibitemOpen
  \bibfield  {author} {\bibinfo {author} {\bibfnamefont {Y.}~\bibnamefont
  {Jun}}, \bibinfo {author} {\bibfnamefont {M.}~\bibnamefont {Gavrilov}}, \
  and\ \bibinfo {author} {\bibfnamefont {J.}~\bibnamefont {Bechhoefer}},\
  }\bibfield  {title} {\enquote {\bibinfo {title} {High-precision test of
  {L}andauer's principle in a feedback trap},}\ }\href@noop {} {\bibfield
  {journal} {\bibinfo  {journal} {Phys. Rev. Lett.}\ }\textbf {\bibinfo
  {volume} {113}},\ \bibinfo {pages} {190601} (\bibinfo {year}
  {2014})}\BibitemShut {NoStop}%
\bibitem [{\citenamefont {B{\'e}rut}\ \emph {et~al.}(2012)\citenamefont
  {B{\'e}rut}, \citenamefont {Arakelyan}, \citenamefont {Petrosyan},
  \citenamefont {Ciliberto}, \citenamefont {Dillenschneider},\ and\
  \citenamefont {Lutz}}]{berut2012experimental}%
  \BibitemOpen
  \bibfield  {author} {\bibinfo {author} {\bibfnamefont {A.}~\bibnamefont
  {B{\'e}rut}}, \bibinfo {author} {\bibfnamefont {A.}~\bibnamefont
  {Arakelyan}}, \bibinfo {author} {\bibfnamefont {A.}~\bibnamefont
  {Petrosyan}}, \bibinfo {author} {\bibfnamefont {S.}~\bibnamefont
  {Ciliberto}}, \bibinfo {author} {\bibfnamefont {R.}~\bibnamefont
  {Dillenschneider}}, \ and\ \bibinfo {author} {\bibfnamefont {E.}~\bibnamefont
  {Lutz}},\ }\bibfield  {title} {\enquote {\bibinfo {title} {Experimental
  verification of {L}andauer's principle linking information and
  thermodynamics},}\ }\href@noop {} {\bibfield  {journal} {\bibinfo  {journal}
  {Nature}\ }\textbf {\bibinfo {volume} {483}},\ \bibinfo {pages} {187}
  (\bibinfo {year} {2012})}\BibitemShut {NoStop}%
\bibitem [{\citenamefont {Toyabe}\ \emph {et~al.}(2010)\citenamefont {Toyabe},
  \citenamefont {Okamoto}, \citenamefont {Watanabe-Nakayama}, \citenamefont
  {Taketani}, \citenamefont {Kudo},\ and\ \citenamefont
  {Muneyuki}}]{toyabe2010nonequilibrium}%
  \BibitemOpen
  \bibfield  {author} {\bibinfo {author} {\bibfnamefont {S.}~\bibnamefont
  {Toyabe}}, \bibinfo {author} {\bibfnamefont {T.}~\bibnamefont {Okamoto}},
  \bibinfo {author} {\bibfnamefont {T.}~\bibnamefont {Watanabe-Nakayama}},
  \bibinfo {author} {\bibfnamefont {H.}~\bibnamefont {Taketani}}, \bibinfo
  {author} {\bibfnamefont {S.}~\bibnamefont {Kudo}}, \ and\ \bibinfo {author}
  {\bibfnamefont {E.}~\bibnamefont {Muneyuki}},\ }\bibfield  {title} {\enquote
  {\bibinfo {title} {Nonequilibrium energetics of a single {F}$_1$-{ATP}ase
  molecule},}\ }\href@noop {} {\bibfield  {journal} {\bibinfo  {journal} {Phys.
  Rev. Lett.}\ }\textbf {\bibinfo {volume} {104}},\ \bibinfo {pages} {198103}
  (\bibinfo {year} {2010})}\BibitemShut {NoStop}%
\bibitem [{\citenamefont {Blickle}\ and\ \citenamefont
  {Bechinger}(2012)}]{blickle2012realization}%
  \BibitemOpen
  \bibfield  {author} {\bibinfo {author} {\bibfnamefont {V.}~\bibnamefont
  {Blickle}}\ and\ \bibinfo {author} {\bibfnamefont {C.}~\bibnamefont
  {Bechinger}},\ }\bibfield  {title} {\enquote {\bibinfo {title} {Realization
  of a micrometre-sized stochastic heat engine},}\ }\href@noop {} {\bibfield
  {journal} {\bibinfo  {journal} {Nat. Phys.}\ }\textbf {\bibinfo {volume}
  {8}},\ \bibinfo {pages} {143--146} (\bibinfo {year} {2012})}\BibitemShut
  {NoStop}%
\bibitem [{\citenamefont {Quinto-Su}(2014)}]{quinto2014microscopic}%
  \BibitemOpen
  \bibfield  {author} {\bibinfo {author} {\bibfnamefont {P.~A.}\ \bibnamefont
  {Quinto-Su}},\ }\bibfield  {title} {\enquote {\bibinfo {title} {A microscopic
  steam engine implemented in an optical tweezer},}\ }\href@noop {} {\bibfield
  {journal} {\bibinfo  {journal} {Nat. Commun.}\ }\textbf {\bibinfo {volume}
  {5}},\ \bibinfo {pages} {5889} (\bibinfo {year} {2014})}\BibitemShut
  {NoStop}%
\bibitem [{\citenamefont {Mart{\'\i}nez}\ \emph {et~al.}(2016)\citenamefont
  {Mart{\'\i}nez}, \citenamefont {Rold{\'a}n}, \citenamefont {Dinis},
  \citenamefont {Petrov}, \citenamefont {Parrondo},\ and\ \citenamefont
  {Rica}}]{martinez2016brownian}%
  \BibitemOpen
  \bibfield  {author} {\bibinfo {author} {\bibfnamefont {I.~A.}\ \bibnamefont
  {Mart{\'\i}nez}}, \bibinfo {author} {\bibfnamefont {E.}~\bibnamefont
  {Rold{\'a}n}}, \bibinfo {author} {\bibfnamefont {L.}~\bibnamefont {Dinis}},
  \bibinfo {author} {\bibfnamefont {D.}~\bibnamefont {Petrov}}, \bibinfo
  {author} {\bibfnamefont {J.~M.~R.}\ \bibnamefont {Parrondo}}, \ and\ \bibinfo
  {author} {\bibfnamefont {R.~A.}\ \bibnamefont {Rica}},\ }\bibfield  {title}
  {\enquote {\bibinfo {title} {Brownian {C}arnot engine},}\ }\href@noop {}
  {\bibfield  {journal} {\bibinfo  {journal} {Nat. Phys.}\ }\textbf {\bibinfo
  {volume} {12}},\ \bibinfo {pages} {67--70} (\bibinfo {year}
  {2016})}\BibitemShut {NoStop}%
\bibitem [{\citenamefont {Mart{\'\i}nez}\ \emph {et~al.}(2017)\citenamefont
  {Mart{\'\i}nez}, \citenamefont {Rold{\'a}n}, \citenamefont {Dinis},\ and\
  \citenamefont {Rica}}]{martinez2017colloidal}%
  \BibitemOpen
  \bibfield  {author} {\bibinfo {author} {\bibfnamefont {I.~A.}\ \bibnamefont
  {Mart{\'\i}nez}}, \bibinfo {author} {\bibfnamefont {E.}~\bibnamefont
  {Rold{\'a}n}}, \bibinfo {author} {\bibfnamefont {L.}~\bibnamefont {Dinis}}, \
  and\ \bibinfo {author} {\bibfnamefont {R.~A.}\ \bibnamefont {Rica}},\
  }\bibfield  {title} {\enquote {\bibinfo {title} {Colloidal heat engines: a
  review},}\ }\href@noop {} {\bibfield  {journal} {\bibinfo  {journal} {Soft
  Matter}\ }\textbf {\bibinfo {volume} {13}},\ \bibinfo {pages} {22--36}
  (\bibinfo {year} {2017})}\BibitemShut {NoStop}%
\bibitem [{\citenamefont {Argun}\ \emph {et~al.}(2017)\citenamefont {Argun},
  \citenamefont {Soni}, \citenamefont {Dabelow}, \citenamefont {Bo},
  \citenamefont {Pesce}, \citenamefont {Eichhorn},\ and\ \citenamefont
  {Volpe}}]{argun2017experimental}%
  \BibitemOpen
  \bibfield  {author} {\bibinfo {author} {\bibfnamefont {A.}~\bibnamefont
  {Argun}}, \bibinfo {author} {\bibfnamefont {J.}~\bibnamefont {Soni}},
  \bibinfo {author} {\bibfnamefont {L.}~\bibnamefont {Dabelow}}, \bibinfo
  {author} {\bibfnamefont {S.}~\bibnamefont {Bo}}, \bibinfo {author}
  {\bibfnamefont {G.}~\bibnamefont {Pesce}}, \bibinfo {author} {\bibfnamefont
  {R.}~\bibnamefont {Eichhorn}}, \ and\ \bibinfo {author} {\bibfnamefont
  {G.}~\bibnamefont {Volpe}},\ }\bibfield  {title} {\enquote {\bibinfo {title}
  {Experimental realization of a minimal microscopic heat engine},}\
  }\href@noop {} {\bibfield  {journal} {\bibinfo  {journal} {Phys. Rev. E}\
  }\textbf {\bibinfo {volume} {96}},\ \bibinfo {pages} {052106} (\bibinfo
  {year} {2017})}\BibitemShut {NoStop}%
\bibitem [{\citenamefont {Schmidt}\ \emph {et~al.}(2018)\citenamefont
  {Schmidt}, \citenamefont {Magazz{\`u}}, \citenamefont {Callegari},
  \citenamefont {Biancofiore}, \citenamefont {Cichos},\ and\ \citenamefont
  {Volpe}}]{schmidt2018microscopic}%
  \BibitemOpen
  \bibfield  {author} {\bibinfo {author} {\bibfnamefont {F.}~\bibnamefont
  {Schmidt}}, \bibinfo {author} {\bibfnamefont {A.}~\bibnamefont
  {Magazz{\`u}}}, \bibinfo {author} {\bibfnamefont {A.}~\bibnamefont
  {Callegari}}, \bibinfo {author} {\bibfnamefont {L.}~\bibnamefont
  {Biancofiore}}, \bibinfo {author} {\bibfnamefont {F.}~\bibnamefont {Cichos}},
  \ and\ \bibinfo {author} {\bibfnamefont {G.}~\bibnamefont {Volpe}},\
  }\bibfield  {title} {\enquote {\bibinfo {title} {Microscopic engine powered
  by critical demixing},}\ }\href@noop {} {\bibfield  {journal} {\bibinfo
  {journal} {Phys. Rev. Lett.}\ }\textbf {\bibinfo {volume} {120}},\ \bibinfo
  {pages} {068004} (\bibinfo {year} {2018})}\BibitemShut {NoStop}%
\bibitem [{\citenamefont {Gavrilov}\ \emph {et~al.}(2014)\citenamefont
  {Gavrilov}, \citenamefont {Jun},\ and\ \citenamefont
  {Bechhoefer}}]{gavrilov2014real}%
  \BibitemOpen
  \bibfield  {author} {\bibinfo {author} {\bibfnamefont {M.}~\bibnamefont
  {Gavrilov}}, \bibinfo {author} {\bibfnamefont {Y.}~\bibnamefont {Jun}}, \
  and\ \bibinfo {author} {\bibfnamefont {J.}~\bibnamefont {Bechhoefer}},\
  }\bibfield  {title} {\enquote {\bibinfo {title} {Real-time calibration of a
  feedback trap},}\ }\href@noop {} {\bibfield  {journal} {\bibinfo  {journal}
  {Rev. Sci. Instrumen.}\ }\textbf {\bibinfo {volume} {85}},\ \bibinfo {pages}
  {095102} (\bibinfo {year} {2014})}\BibitemShut {NoStop}%
\bibitem [{\citenamefont {Wu}\ \emph {et~al.}(2009)\citenamefont {Wu},
  \citenamefont {Huang}, \citenamefont {Tischer}, \citenamefont {Jonas},\ and\
  \citenamefont {Florin}}]{wu2009direct}%
  \BibitemOpen
  \bibfield  {author} {\bibinfo {author} {\bibfnamefont {P.}~\bibnamefont
  {Wu}}, \bibinfo {author} {\bibfnamefont {R.}~\bibnamefont {Huang}}, \bibinfo
  {author} {\bibfnamefont {C.}~\bibnamefont {Tischer}}, \bibinfo {author}
  {\bibfnamefont {A.}~\bibnamefont {Jonas}}, \ and\ \bibinfo {author}
  {\bibfnamefont {E.-L.}\ \bibnamefont {Florin}},\ }\bibfield  {title}
  {\enquote {\bibinfo {title} {Direct measurement of the nonconservative force
  field generated by optical tweezers},}\ }\href@noop {} {\bibfield  {journal}
  {\bibinfo  {journal} {Phys. Rev. Lett.}\ }\textbf {\bibinfo {volume} {103}},\
  \bibinfo {pages} {108101} (\bibinfo {year} {2009})}\BibitemShut {NoStop}%
\bibitem [{\citenamefont {Friedrich}\ \emph {et~al.}(2011)\citenamefont
  {Friedrich}, \citenamefont {Peinke}, \citenamefont {Sahimi},\ and\
  \citenamefont {Tabar}}]{friedrich2011approaching}%
  \BibitemOpen
  \bibfield  {author} {\bibinfo {author} {\bibfnamefont {R.}~\bibnamefont
  {Friedrich}}, \bibinfo {author} {\bibfnamefont {J.}~\bibnamefont {Peinke}},
  \bibinfo {author} {\bibfnamefont {M.}~\bibnamefont {Sahimi}}, \ and\ \bibinfo
  {author} {\bibfnamefont {M~Reza~Rahimi}\ \bibnamefont {Tabar}},\ }\bibfield
  {title} {\enquote {\bibinfo {title} {Approaching complexity by stochastic
  methods: From biological systems to turbulence},}\ }\href@noop {} {\bibfield
  {journal} {\bibinfo  {journal} {Phys. Rep.}\ }\textbf {\bibinfo {volume}
  {506}},\ \bibinfo {pages} {87--162} (\bibinfo {year} {2011})}\BibitemShut
  {NoStop}%
\bibitem [{\citenamefont {Ciliberto}(2017)}]{ciliberto2017experiments}%
  \BibitemOpen
  \bibfield  {author} {\bibinfo {author} {\bibfnamefont {S.}~\bibnamefont
  {Ciliberto}},\ }\bibfield  {title} {\enquote {\bibinfo {title} {Experiments
  in stochastic thermodynamics: Short history and perspectives},}\ }\href@noop
  {} {\bibfield  {journal} {\bibinfo  {journal} {Phys. Rev. X}\ }\textbf
  {\bibinfo {volume} {7}},\ \bibinfo {pages} {021051} (\bibinfo {year}
  {2017})}\BibitemShut {NoStop}%
\bibitem [{\citenamefont {Berg-S{\o}rensen}\ and\ \citenamefont
  {Flyvbjerg}(2004)}]{berg2004power}%
  \BibitemOpen
  \bibfield  {author} {\bibinfo {author} {\bibfnamefont {K.}~\bibnamefont
  {Berg-S{\o}rensen}}\ and\ \bibinfo {author} {\bibfnamefont {H.}~\bibnamefont
  {Flyvbjerg}},\ }\bibfield  {title} {\enquote {\bibinfo {title} {Power
  spectrum analysis for optical tweezers},}\ }\href@noop {} {\bibfield
  {journal} {\bibinfo  {journal} {Rev. Sci. Instrumen.}\ }\textbf {\bibinfo
  {volume} {75}},\ \bibinfo {pages} {594--612} (\bibinfo {year}
  {2004})}\BibitemShut {NoStop}%
\bibitem [{\citenamefont {Garc{\'\i}a}\ \emph {et~al.}(2018)\citenamefont
  {Garc{\'\i}a}, \citenamefont {P{\'e}rez}, \citenamefont {Volpe},
  \citenamefont {Arzola},\ and\ \citenamefont {Volpe}}]{garcia2018high}%
  \BibitemOpen
  \bibfield  {author} {\bibinfo {author} {\bibfnamefont {L.~P.}\ \bibnamefont
  {Garc{\'\i}a}}, \bibinfo {author} {\bibfnamefont {J.~D.}\ \bibnamefont
  {P{\'e}rez}}, \bibinfo {author} {\bibfnamefont {G.}~\bibnamefont {Volpe}},
  \bibinfo {author} {\bibfnamefont {A.~V.}\ \bibnamefont {Arzola}}, \ and\
  \bibinfo {author} {\bibfnamefont {G.}~\bibnamefont {Volpe}},\ }\bibfield
  {title} {\enquote {\bibinfo {title} {High-performance reconstruction of
  microscopic force fields from {B}rownian trajectories},}\ }\href@noop {}
  {\bibfield  {journal} {\bibinfo  {journal} {Nat. Commun.}\ }\textbf {\bibinfo
  {volume} {9}},\ \bibinfo {pages} {5166} (\bibinfo {year} {2018})}\BibitemShut
  {NoStop}%
\bibitem [{\citenamefont {B{\"o}ttcher}\ \emph {et~al.}(2006)\citenamefont
  {B{\"o}ttcher}, \citenamefont {Peinke}, \citenamefont {Kleinhans},
  \citenamefont {Friedrich}, \citenamefont {Lind},\ and\ \citenamefont
  {Haase}}]{bottcher2006reconstruction}%
  \BibitemOpen
  \bibfield  {author} {\bibinfo {author} {\bibfnamefont {F.}~\bibnamefont
  {B{\"o}ttcher}}, \bibinfo {author} {\bibfnamefont {J.}~\bibnamefont
  {Peinke}}, \bibinfo {author} {\bibfnamefont {D.}~\bibnamefont {Kleinhans}},
  \bibinfo {author} {\bibfnamefont {R.}~\bibnamefont {Friedrich}}, \bibinfo
  {author} {\bibfnamefont {P.~G.}\ \bibnamefont {Lind}}, \ and\ \bibinfo
  {author} {\bibfnamefont {M.}~\bibnamefont {Haase}},\ }\bibfield  {title}
  {\enquote {\bibinfo {title} {Reconstruction of complex dynamical systems
  affected by strong measurement noise},}\ }\href@noop {} {\bibfield  {journal}
  {\bibinfo  {journal} {Phys. Rev. Lett.}\ }\textbf {\bibinfo {volume} {97}},\
  \bibinfo {pages} {090603} (\bibinfo {year} {2006})}\BibitemShut {NoStop}%
\bibitem [{\citenamefont {T{\"u}rkcan}\ \emph {et~al.}(2012)\citenamefont
  {T{\"u}rkcan}, \citenamefont {Alexandrou},\ and\ \citenamefont
  {Masson}}]{turkcan2012bayesian}%
  \BibitemOpen
  \bibfield  {author} {\bibinfo {author} {\bibfnamefont {S.}~\bibnamefont
  {T{\"u}rkcan}}, \bibinfo {author} {\bibfnamefont {A.}~\bibnamefont
  {Alexandrou}}, \ and\ \bibinfo {author} {\bibfnamefont {J.~B.}\ \bibnamefont
  {Masson}},\ }\bibfield  {title} {\enquote {\bibinfo {title} {A {B}ayesian
  inference scheme to extract diffusivity and potential fields from confined
  single-molecule trajectories},}\ }\href@noop {} {\bibfield  {journal}
  {\bibinfo  {journal} {Biophys. J.}\ }\textbf {\bibinfo {volume} {102}},\
  \bibinfo {pages} {2288--2298} (\bibinfo {year} {2012})}\BibitemShut {NoStop}%
\bibitem [{\citenamefont {Bera}\ \emph {et~al.}(2017)\citenamefont {Bera},
  \citenamefont {Paul}, \citenamefont {Singh}, \citenamefont {Ghosh},
  \citenamefont {Kundu}, \citenamefont {Banerjee},\ and\ \citenamefont
  {Adhikari}}]{bera2017fast}%
  \BibitemOpen
  \bibfield  {author} {\bibinfo {author} {\bibfnamefont {S.}~\bibnamefont
  {Bera}}, \bibinfo {author} {\bibfnamefont {S.}~\bibnamefont {Paul}}, \bibinfo
  {author} {\bibfnamefont {R.}~\bibnamefont {Singh}}, \bibinfo {author}
  {\bibfnamefont {D.}~\bibnamefont {Ghosh}}, \bibinfo {author} {\bibfnamefont
  {A.}~\bibnamefont {Kundu}}, \bibinfo {author} {\bibfnamefont
  {A.}~\bibnamefont {Banerjee}}, \ and\ \bibinfo {author} {\bibfnamefont
  {R.}~\bibnamefont {Adhikari}},\ }\bibfield  {title} {\enquote {\bibinfo
  {title} {Fast {B}ayesian inference of optical trap stiffness and particle
  diffusion},}\ }\href@noop {} {\bibfield  {journal} {\bibinfo  {journal} {Sci.
  Rep.}\ }\textbf {\bibinfo {volume} {7}},\ \bibinfo {pages} {1--10} (\bibinfo
  {year} {2017})}\BibitemShut {NoStop}%
\bibitem [{\citenamefont {Frishman}\ and\ \citenamefont
  {Ronceray}(2020)}]{frishman2020learning}%
  \BibitemOpen
  \bibfield  {author} {\bibinfo {author} {\bibfnamefont {A.}~\bibnamefont
  {Frishman}}\ and\ \bibinfo {author} {\bibfnamefont {P.}~\bibnamefont
  {Ronceray}},\ }\bibfield  {title} {\enquote {\bibinfo {title} {Learning force
  fields from stochastic trajectories},}\ }\href@noop {} {\bibfield  {journal}
  {\bibinfo  {journal} {Phys. Rev. X}\ }\textbf {\bibinfo {volume} {10}},\
  \bibinfo {pages} {021009} (\bibinfo {year} {2020})}\BibitemShut {NoStop}%
\bibitem [{\citenamefont {Lipton}\ \emph {et~al.}(2015)\citenamefont {Lipton},
  \citenamefont {Berkowitz},\ and\ \citenamefont {Elkan}}]{lipton2015critical}%
  \BibitemOpen
  \bibfield  {author} {\bibinfo {author} {\bibfnamefont {Z.~C.}\ \bibnamefont
  {Lipton}}, \bibinfo {author} {\bibfnamefont {J.}~\bibnamefont {Berkowitz}}, \
  and\ \bibinfo {author} {\bibfnamefont {C.}~\bibnamefont {Elkan}},\ }\bibfield
   {title} {\enquote {\bibinfo {title} {A critical review of recurrent neural
  networks for sequence learning},}\ }\href@noop {} {\bibfield  {journal}
  {\bibinfo  {journal} {arXiv preprint arXiv:1506.00019}\ } (\bibinfo {year}
  {2015})}\BibitemShut {NoStop}%
\bibitem [{\citenamefont {Graves}\ \emph {et~al.}(2013)\citenamefont {Graves},
  \citenamefont {Jaitly},\ and\ \citenamefont {Mohamed}}]{graves2013hybrid}%
  \BibitemOpen
  \bibfield  {author} {\bibinfo {author} {\bibfnamefont {A.}~\bibnamefont
  {Graves}}, \bibinfo {author} {\bibfnamefont {N.}~\bibnamefont {Jaitly}}, \
  and\ \bibinfo {author} {\bibfnamefont {A.}~\bibnamefont {Mohamed}},\
  }\bibfield  {title} {\enquote {\bibinfo {title} {Hybrid speech recognition
  with deep bidirectional lstm},}\ }in\ \href@noop {} {\emph {\bibinfo
  {booktitle} {2013 IEEE workshop on automatic speech recognition and
  understanding}}}\ (\bibinfo {organization} {IEEE},\ \bibinfo {year} {2013})\
  pp.\ \bibinfo {pages} {273--278}\BibitemShut {NoStop}%
\bibitem [{\citenamefont {Wu}\ \emph {et~al.}(2016)\citenamefont {Wu},
  \citenamefont {Schuster}, \citenamefont {Chen}, \citenamefont {Le},
  \citenamefont {Norouzi}, \citenamefont {Macherey}, \citenamefont {Krikun},
  \citenamefont {Cao}, \citenamefont {Gao}, \citenamefont {Macherey} \emph
  {et~al.}}]{wu2016google}%
  \BibitemOpen
  \bibfield  {author} {\bibinfo {author} {\bibfnamefont {Y.}~\bibnamefont
  {Wu}}, \bibinfo {author} {\bibfnamefont {M.}~\bibnamefont {Schuster}},
  \bibinfo {author} {\bibfnamefont {Z.}~\bibnamefont {Chen}}, \bibinfo {author}
  {\bibfnamefont {Q.~V.}\ \bibnamefont {Le}}, \bibinfo {author} {\bibfnamefont
  {M.}~\bibnamefont {Norouzi}}, \bibinfo {author} {\bibfnamefont
  {W.}~\bibnamefont {Macherey}}, \bibinfo {author} {\bibfnamefont
  {M.}~\bibnamefont {Krikun}}, \bibinfo {author} {\bibfnamefont
  {Y.}~\bibnamefont {Cao}}, \bibinfo {author} {\bibfnamefont {Q.}~\bibnamefont
  {Gao}}, \bibinfo {author} {\bibfnamefont {K.}~\bibnamefont {Macherey}},
  \emph {et~al.},\ }\bibfield  {title} {\enquote {\bibinfo {title} {Google's
  neural machine translation system: Bridging the gap between human and machine
  translation},}\ }\href@noop {} {\bibfield  {journal} {\bibinfo  {journal}
  {arXiv preprint arXiv:1609.08144}\ } (\bibinfo {year} {2016})}\BibitemShut
  {NoStop}%
\bibitem [{\citenamefont {Han}\ \emph {et~al.}(2017)\citenamefont {Han},
  \citenamefont {Kang}, \citenamefont {Mao}, \citenamefont {Hu}, \citenamefont
  {Li}, \citenamefont {Li}, \citenamefont {Xie}, \citenamefont {Luo},
  \citenamefont {Yao}, \citenamefont {Wang} \emph {et~al.}}]{han2017ese}%
  \BibitemOpen
  \bibfield  {author} {\bibinfo {author} {\bibfnamefont {S.}~\bibnamefont
  {Han}}, \bibinfo {author} {\bibfnamefont {J.}~\bibnamefont {Kang}}, \bibinfo
  {author} {\bibfnamefont {H.}~\bibnamefont {Mao}}, \bibinfo {author}
  {\bibfnamefont {Y.}~\bibnamefont {Hu}}, \bibinfo {author} {\bibfnamefont
  {X.}~\bibnamefont {Li}}, \bibinfo {author} {\bibfnamefont {Y.}~\bibnamefont
  {Li}}, \bibinfo {author} {\bibfnamefont {D.}~\bibnamefont {Xie}}, \bibinfo
  {author} {\bibfnamefont {H.}~\bibnamefont {Luo}}, \bibinfo {author}
  {\bibfnamefont {S.}~\bibnamefont {Yao}}, \bibinfo {author} {\bibfnamefont
  {Y.}~\bibnamefont {Wang}},  \emph {et~al.},\ }\bibfield  {title} {\enquote
  {\bibinfo {title} {Ese: Efficient speech recognition engine with sparse lstm
  on fpga},}\ }in\ \href@noop {} {\emph {\bibinfo {booktitle} {Proceedings of
  the 2017 ACM/SIGDA International Symposium on Field-Programmable Gate
  Arrays}}}\ (\bibinfo {year} {2017})\ pp.\ \bibinfo {pages}
  {75--84}\BibitemShut {NoStop}%
\bibitem [{\citenamefont {Gers}\ \emph {et~al.}(1999)\citenamefont {Gers},
  \citenamefont {Schmidhuber},\ and\ \citenamefont
  {Cummins}}]{gers1999learning}%
  \BibitemOpen
  \bibfield  {author} {\bibinfo {author} {\bibfnamefont {F.~A.}\ \bibnamefont
  {Gers}}, \bibinfo {author} {\bibfnamefont {J.}~\bibnamefont {Schmidhuber}}, \
  and\ \bibinfo {author} {\bibfnamefont {Fred}\ \bibnamefont {Cummins}},\
  }\bibfield  {title} {\enquote {\bibinfo {title} {Learning to forget:
  Continual prediction with {LSTM}},}\ }\href@noop {} {\bibfield  {journal}
  {\bibinfo  {journal} {IET Conf. Proc.}\ ,\ \bibinfo {pages} {850--855}}
  (\bibinfo {year} {1999})}\BibitemShut {NoStop}%
\bibitem [{\citenamefont {Bo}\ \emph {et~al.}(2019)\citenamefont {Bo},
  \citenamefont {Schmidt}, \citenamefont {Eichhorn},\ and\ \citenamefont
  {Volpe}}]{bo2019measurement}%
  \BibitemOpen
  \bibfield  {author} {\bibinfo {author} {\bibfnamefont {S.}~\bibnamefont
  {Bo}}, \bibinfo {author} {\bibfnamefont {F.}~\bibnamefont {Schmidt}},
  \bibinfo {author} {\bibfnamefont {R.}~\bibnamefont {Eichhorn}}, \ and\
  \bibinfo {author} {\bibfnamefont {G.}~\bibnamefont {Volpe}},\ }\bibfield
  {title} {\enquote {\bibinfo {title} {Measurement of anomalous diffusion using
  recurrent neural networks},}\ }\href@noop {} {\bibfield  {journal} {\bibinfo
  {journal} {Phys. Rev. E}\ }\textbf {\bibinfo {volume} {100}},\ \bibinfo
  {pages} {010102} (\bibinfo {year} {2019})}\BibitemShut {NoStop}%
\bibitem [{\citenamefont {Argun}\ \emph {et~al.}(2020)\citenamefont {Argun},
  \citenamefont {Thalheim}, \citenamefont {Bo}, \citenamefont {Cichos},\ and\
  \citenamefont {Volpe}}]{DC}%
  \BibitemOpen
  \bibfield  {author} {\bibinfo {author} {\bibfnamefont {A.}~\bibnamefont
  {Argun}}, \bibinfo {author} {\bibfnamefont {T.}~\bibnamefont {Thalheim}},
  \bibinfo {author} {\bibfnamefont {S.}~\bibnamefont {Bo}}, \bibinfo {author}
  {\bibfnamefont {F.}~\bibnamefont {Cichos}}, \ and\ \bibinfo {author}
  {\bibfnamefont {G.}~\bibnamefont {Volpe}},\ }\href
  {github.com/softmatterlab/DeepCalib} {\enquote {\bibinfo {title}
  {{DeepCalib}},}\ }\bibinfo {howpublished}
  {http://github.com/softmatterlab/DeepCalib} (\bibinfo {year}
  {2020})\BibitemShut {NoStop}%
\bibitem [{\citenamefont {Zdeborov{\'a}}(2017)}]{zdeborova2017machine}%
  \BibitemOpen
  \bibfield  {author} {\bibinfo {author} {\bibfnamefont {L.}~\bibnamefont
  {Zdeborov{\'a}}},\ }\bibfield  {title} {\enquote {\bibinfo {title} {Machine
  learning: New tool in the box},}\ }\href@noop {} {\bibfield  {journal}
  {\bibinfo  {journal} {Nat. Phys.}\ }\textbf {\bibinfo {volume} {13}},\
  \bibinfo {pages} {420--421} (\bibinfo {year} {2017})}\BibitemShut {NoStop}%
\bibitem [{\citenamefont {Cichos}\ \emph {et~al.}(2020)\citenamefont {Cichos},
  \citenamefont {Gustavsson}, \citenamefont {Mehlig},\ and\ \citenamefont
  {Volpe}}]{cichos2020machine}%
  \BibitemOpen
  \bibfield  {author} {\bibinfo {author} {\bibfnamefont {F.}~\bibnamefont
  {Cichos}}, \bibinfo {author} {\bibfnamefont {K.}~\bibnamefont {Gustavsson}},
  \bibinfo {author} {\bibfnamefont {B.}~\bibnamefont {Mehlig}}, \ and\ \bibinfo
  {author} {\bibfnamefont {G.}~\bibnamefont {Volpe}},\ }\bibfield  {title}
  {\enquote {\bibinfo {title} {Machine learning for active matter},}\
  }\href@noop {} {\bibfield  {journal} {\bibinfo  {journal} {Nat. Mach.
  Intell.}\ }\textbf {\bibinfo {volume} {2}},\ \bibinfo {pages} {94--103}
  (\bibinfo {year} {2020})}\BibitemShut {NoStop}%
\bibitem [{\citenamefont {Nielsen}(2015)}]{nielsen2015neural}%
  \BibitemOpen
  \bibfield  {author} {\bibinfo {author} {\bibfnamefont {M.~A.}\ \bibnamefont
  {Nielsen}},\ }\href@noop {} {\emph {\bibinfo {title} {Neural networks and
  deep learning}}},\ Vol.\ \bibinfo {volume} {2018}\ (\bibinfo  {publisher}
  {Determination press San Francisco, CA, USA:},\ \bibinfo {year}
  {2015})\BibitemShut {NoStop}%
\bibitem [{\citenamefont {Chollet}\ \emph {et~al.}(2018)\citenamefont {Chollet}
  \emph {et~al.}}]{chollet2018keras}%
  \BibitemOpen
  \bibfield  {author} {\bibinfo {author} {\bibfnamefont {Fran{\c{c}}ois}\
  \bibnamefont {Chollet}} \emph {et~al.},\ }\bibfield  {title} {\enquote
  {\bibinfo {title} {Keras: The {P}ython deep learning library},}\ }\href@noop
  {} {\bibfield  {journal} {\bibinfo  {journal} {Astrophysics Source Code
  Library}\ } (\bibinfo {year} {2018})}\BibitemShut {NoStop}%
\bibitem [{\citenamefont {McClelland}\ \emph {et~al.}(1986)\citenamefont
  {McClelland}, \citenamefont {Rumelhart}, \citenamefont {Group} \emph
  {et~al.}}]{mcclelland1986parallel}%
  \BibitemOpen
  \bibfield  {author} {\bibinfo {author} {\bibfnamefont {J.~L.}\ \bibnamefont
  {McClelland}}, \bibinfo {author} {\bibfnamefont {D.~E.}\ \bibnamefont
  {Rumelhart}}, \bibinfo {author} {\bibfnamefont {PDP~Research}\ \bibnamefont
  {Group}},  \emph {et~al.},\ }\href@noop {} {\emph {\bibinfo {title} {Parallel
  distributed processing: Explorations in the Microstructure of Cognition}}}\
  (\bibinfo  {publisher} {MIT Press Cambridge},\ \bibinfo {year}
  {1986})\BibitemShut {NoStop}%
\bibitem [{\citenamefont {Granik}\ \emph {et~al.}(2019)\citenamefont {Granik},
  \citenamefont {Weiss}, \citenamefont {Nehme}, \citenamefont {Levin},
  \citenamefont {Chein}, \citenamefont {Perlson}, \citenamefont {Roichman},\
  and\ \citenamefont {Shechtman}}]{granik2019single}%
  \BibitemOpen
  \bibfield  {author} {\bibinfo {author} {\bibfnamefont {N.}~\bibnamefont
  {Granik}}, \bibinfo {author} {\bibfnamefont {L.~E.}\ \bibnamefont {Weiss}},
  \bibinfo {author} {\bibfnamefont {E.}~\bibnamefont {Nehme}}, \bibinfo
  {author} {\bibfnamefont {M.}~\bibnamefont {Levin}}, \bibinfo {author}
  {\bibfnamefont {M.}~\bibnamefont {Chein}}, \bibinfo {author} {\bibfnamefont
  {E.}~\bibnamefont {Perlson}}, \bibinfo {author} {\bibfnamefont
  {Y.}~\bibnamefont {Roichman}}, \ and\ \bibinfo {author} {\bibfnamefont
  {Y.}~\bibnamefont {Shechtman}},\ }\bibfield  {title} {\enquote {\bibinfo
  {title} {Single-particle diffusion characterization by deep learning},}\
  }\href@noop {} {\bibfield  {journal} {\bibinfo  {journal} {Biophys. J.}\
  }\textbf {\bibinfo {volume} {117}},\ \bibinfo {pages} {185--192} (\bibinfo
  {year} {2019})}\BibitemShut {NoStop}%
\bibitem [{\citenamefont {Mu{\~n}oz-Gil}\ \emph {et~al.}(2020)\citenamefont
  {Mu{\~n}oz-Gil}, \citenamefont {Garcia-March}, \citenamefont {Manzo},
  \citenamefont {Mart{\'\i}n-Guerrero},\ and\ \citenamefont
  {Lewenstein}}]{munoz2020single}%
  \BibitemOpen
  \bibfield  {author} {\bibinfo {author} {\bibfnamefont {G.}~\bibnamefont
  {Mu{\~n}oz-Gil}}, \bibinfo {author} {\bibfnamefont {M.~A.}\ \bibnamefont
  {Garcia-March}}, \bibinfo {author} {\bibfnamefont {C.}~\bibnamefont {Manzo}},
  \bibinfo {author} {\bibfnamefont {J.~D.}\ \bibnamefont
  {Mart{\'\i}n-Guerrero}}, \ and\ \bibinfo {author} {\bibfnamefont
  {M.}~\bibnamefont {Lewenstein}},\ }\bibfield  {title} {\enquote {\bibinfo
  {title} {Single trajectory characterization via machine learning},}\
  }\href@noop {} {\bibfield  {journal} {\bibinfo  {journal} {New J. Phys.}\
  }\textbf {\bibinfo {volume} {22}},\ \bibinfo {pages} {013010} (\bibinfo
  {year} {2020})}\BibitemShut {NoStop}%
\bibitem [{\citenamefont {Seif}\ \emph {et~al.}(2019)\citenamefont {Seif},
  \citenamefont {Hafezi},\ and\ \citenamefont {Jarzynski}}]{seif2019machine}%
  \BibitemOpen
  \bibfield  {author} {\bibinfo {author} {\bibfnamefont {A.}~\bibnamefont
  {Seif}}, \bibinfo {author} {\bibfnamefont {M.}~\bibnamefont {Hafezi}}, \ and\
  \bibinfo {author} {\bibfnamefont {C.}~\bibnamefont {Jarzynski}},\ }\bibfield
  {title} {\enquote {\bibinfo {title} {Machine learning the thermodynamic arrow
  of time},}\ }\href@noop {} {\bibfield  {journal} {\bibinfo  {journal} {arXiv
  preprint arXiv:1909.12380}\ } (\bibinfo {year} {2019})}\BibitemShut {NoStop}%
\bibitem [{\citenamefont {Hannel}\ \emph {et~al.}(2018)\citenamefont {Hannel},
  \citenamefont {Abdulali}, \citenamefont {O'Brien},\ and\ \citenamefont
  {Grier}}]{hannel2018machine}%
  \BibitemOpen
  \bibfield  {author} {\bibinfo {author} {\bibfnamefont {M.~D.}\ \bibnamefont
  {Hannel}}, \bibinfo {author} {\bibfnamefont {A.}~\bibnamefont {Abdulali}},
  \bibinfo {author} {\bibfnamefont {M.}~\bibnamefont {O'Brien}}, \ and\
  \bibinfo {author} {\bibfnamefont {D.~G.}\ \bibnamefont {Grier}},\ }\bibfield
  {title} {\enquote {\bibinfo {title} {Machine-learning techniques for fast and
  accurate feature localization in holograms of colloidal particles},}\
  }\href@noop {} {\bibfield  {journal} {\bibinfo  {journal} {Opt. Express}\
  }\textbf {\bibinfo {volume} {26}},\ \bibinfo {pages} {15221--15231} (\bibinfo
  {year} {2018})}\BibitemShut {NoStop}%
\bibitem [{\citenamefont {Helgadottir}\ \emph {et~al.}(2019)\citenamefont
  {Helgadottir}, \citenamefont {Argun},\ and\ \citenamefont
  {Volpe}}]{helgadottir2019digital}%
  \BibitemOpen
  \bibfield  {author} {\bibinfo {author} {\bibfnamefont {S.}~\bibnamefont
  {Helgadottir}}, \bibinfo {author} {\bibfnamefont {A.}~\bibnamefont {Argun}},
  \ and\ \bibinfo {author} {\bibfnamefont {G.}~\bibnamefont {Volpe}},\
  }\bibfield  {title} {\enquote {\bibinfo {title} {Digital video microscopy
  enhanced by deep learning},}\ }\href@noop {} {\bibfield  {journal} {\bibinfo
  {journal} {Optica}\ }\textbf {\bibinfo {volume} {6}},\ \bibinfo {pages}
  {506--513} (\bibinfo {year} {2019})}\BibitemShut {NoStop}%
\bibitem [{\citenamefont {Barbastathis}\ \emph {et~al.}(2019)\citenamefont
  {Barbastathis}, \citenamefont {Ozcan},\ and\ \citenamefont
  {Situ}}]{barbastathis2019use}%
  \BibitemOpen
  \bibfield  {author} {\bibinfo {author} {\bibfnamefont {G.}~\bibnamefont
  {Barbastathis}}, \bibinfo {author} {\bibfnamefont {A.}~\bibnamefont {Ozcan}},
  \ and\ \bibinfo {author} {\bibfnamefont {G.}~\bibnamefont {Situ}},\
  }\bibfield  {title} {\enquote {\bibinfo {title} {On the use of deep learning
  for computational imaging},}\ }\href@noop {} {\bibfield  {journal} {\bibinfo
  {journal} {Optica}\ }\textbf {\bibinfo {volume} {6}},\ \bibinfo {pages}
  {921--943} (\bibinfo {year} {2019})}\BibitemShut {NoStop}%
\bibitem [{\citenamefont {Gibson}\ \emph {et~al.}(2019)\citenamefont {Gibson},
  \citenamefont {Zhang}, \citenamefont {Stilgoe}, \citenamefont {Nieminen},\
  and\ \citenamefont {Rubinsztein-Dunlop}}]{gibson2019machine}%
  \BibitemOpen
  \bibfield  {author} {\bibinfo {author} {\bibfnamefont {L.~J.}\ \bibnamefont
  {Gibson}}, \bibinfo {author} {\bibfnamefont {S.}~\bibnamefont {Zhang}},
  \bibinfo {author} {\bibfnamefont {A.~B.}\ \bibnamefont {Stilgoe}}, \bibinfo
  {author} {\bibfnamefont {T.~A.}\ \bibnamefont {Nieminen}}, \ and\ \bibinfo
  {author} {\bibfnamefont {H.}~\bibnamefont {Rubinsztein-Dunlop}},\ }\bibfield
  {title} {\enquote {\bibinfo {title} {Machine learning wall effects of
  eccentric spheres for convenient computation},}\ }\href@noop {} {\bibfield
  {journal} {\bibinfo  {journal} {Phys. Rev. E}\ }\textbf {\bibinfo {volume}
  {99}},\ \bibinfo {pages} {043304} (\bibinfo {year} {2019})}\BibitemShut
  {NoStop}%
\bibitem [{\citenamefont {Lenton}\ \emph {et~al.}(2020)\citenamefont {Lenton},
  \citenamefont {Volpe}, \citenamefont {Stilgoe}, \citenamefont {Nieminen},\
  and\ \citenamefont {Rubinsztein-Dunlop}}]{lenton2020machine}%
  \BibitemOpen
  \bibfield  {author} {\bibinfo {author} {\bibfnamefont {I.~C.~D.}\
  \bibnamefont {Lenton}}, \bibinfo {author} {\bibfnamefont {G.}~\bibnamefont
  {Volpe}}, \bibinfo {author} {\bibfnamefont {A.~B.}\ \bibnamefont {Stilgoe}},
  \bibinfo {author} {\bibfnamefont {T.~A}\ \bibnamefont {Nieminen}}, \ and\
  \bibinfo {author} {\bibfnamefont {H.}~\bibnamefont {Rubinsztein-Dunlop}},\
  }\bibfield  {title} {\enquote {\bibinfo {title} {Machine learning reveals
  complex behaviours in optically trapped particles},}\ }\href@noop {}
  {\bibfield  {journal} {\bibinfo  {journal} {arXiv preprint arXiv:2004.08264}\
  } (\bibinfo {year} {2020})}\BibitemShut {NoStop}%
\bibitem [{\citenamefont {Braun}\ \emph {et~al.}(2015)\citenamefont {Braun},
  \citenamefont {Bregulla}, \citenamefont {Günther}, \citenamefont {Mertig},\
  and\ \citenamefont {Cichos}}]{braun2015single}%
  \BibitemOpen
  \bibfield  {author} {\bibinfo {author} {\bibfnamefont {M.}~\bibnamefont
  {Braun}}, \bibinfo {author} {\bibfnamefont {A.~P.}\ \bibnamefont {Bregulla}},
  \bibinfo {author} {\bibfnamefont {K.}~\bibnamefont {Günther}}, \bibinfo
  {author} {\bibfnamefont {M.}~\bibnamefont {Mertig}}, \ and\ \bibinfo {author}
  {\bibfnamefont {F.}~\bibnamefont {Cichos}},\ }\bibfield  {title} {\enquote
  {\bibinfo {title} {Single molecules trapped by dynamic inhomogeneous
  temperature fields},}\ }\href@noop {} {\bibfield  {journal} {\bibinfo
  {journal} {Nano Lett.}\ }\textbf {\bibinfo {volume} {15}},\ \bibinfo {pages}
  {5499--5505} (\bibinfo {year} {2015})}\BibitemShut {NoStop}%
\bibitem [{\citenamefont {Volpe}\ and\ \citenamefont
  {Volpe}(2013)}]{volpe2013simulation}%
  \BibitemOpen
  \bibfield  {author} {\bibinfo {author} {\bibfnamefont {G.}~\bibnamefont
  {Volpe}}\ and\ \bibinfo {author} {\bibfnamefont {G.}~\bibnamefont {Volpe}},\
  }\bibfield  {title} {\enquote {\bibinfo {title} {Simulation of a {B}rownian
  particle in an optical trap},}\ }\href@noop {} {\bibfield  {journal}
  {\bibinfo  {journal} {Am. J. Phys.}\ }\textbf {\bibinfo {volume} {81}},\
  \bibinfo {pages} {224--231} (\bibinfo {year} {2013})}\BibitemShut {NoStop}%
\bibitem [{\citenamefont {Smith}\ \emph {et~al.}(2017)\citenamefont {Smith},
  \citenamefont {Kindermans}, \citenamefont {Ying},\ and\ \citenamefont
  {Le}}]{smith2017don}%
  \BibitemOpen
  \bibfield  {author} {\bibinfo {author} {\bibfnamefont {S.~L.}\ \bibnamefont
  {Smith}}, \bibinfo {author} {\bibfnamefont {P.~J.}\ \bibnamefont
  {Kindermans}}, \bibinfo {author} {\bibfnamefont {C.}~\bibnamefont {Ying}}, \
  and\ \bibinfo {author} {\bibfnamefont {Q.~V.}\ \bibnamefont {Le}},\
  }\bibfield  {title} {\enquote {\bibinfo {title} {Don't decay the learning
  rate, increase the batch size},}\ }\href@noop {} {\bibfield  {journal}
  {\bibinfo  {journal} {arXiv preprint arXiv:1711.00489}\ } (\bibinfo {year}
  {2017})}\BibitemShut {NoStop}%
\bibitem [{\citenamefont {Fr{\"a}nzl}\ \emph {et~al.}(2019)\citenamefont
  {Fr{\"a}nzl}, \citenamefont {Thalheim}, \citenamefont {Adler}, \citenamefont
  {Huster}, \citenamefont {Posseckardt}, \citenamefont {Mertig},\ and\
  \citenamefont {Cichos}}]{franzl2019thermophoretic}%
  \BibitemOpen
  \bibfield  {author} {\bibinfo {author} {\bibfnamefont {M.}~\bibnamefont
  {Fr{\"a}nzl}}, \bibinfo {author} {\bibfnamefont {T.}~\bibnamefont
  {Thalheim}}, \bibinfo {author} {\bibfnamefont {J.}~\bibnamefont {Adler}},
  \bibinfo {author} {\bibfnamefont {D.}~\bibnamefont {Huster}}, \bibinfo
  {author} {\bibfnamefont {J.}~\bibnamefont {Posseckardt}}, \bibinfo {author}
  {\bibfnamefont {M.}~\bibnamefont {Mertig}}, \ and\ \bibinfo {author}
  {\bibfnamefont {F.}~\bibnamefont {Cichos}},\ }\bibfield  {title} {\enquote
  {\bibinfo {title} {Thermophoretic trap for single amyloid fibril and protein
  aggregation studies},}\ }\href@noop {} {\bibfield  {journal} {\bibinfo
  {journal} {Nat. Methods}\ }\textbf {\bibinfo {volume} {16}},\ \bibinfo
  {pages} {611--614} (\bibinfo {year} {2019})}\BibitemShut {NoStop}%
\bibitem [{\citenamefont {W{\"u}rger}(2010)}]{wurger2010thermal}%
  \BibitemOpen
  \bibfield  {author} {\bibinfo {author} {\bibfnamefont {A.}~\bibnamefont
  {W{\"u}rger}},\ }\bibfield  {title} {\enquote {\bibinfo {title} {Thermal
  non-equilibrium transport in colloids},}\ }\href@noop {} {\bibfield
  {journal} {\bibinfo  {journal} {Rep. Prog. Phys.}\ }\textbf {\bibinfo
  {volume} {73}},\ \bibinfo {pages} {126601} (\bibinfo {year}
  {2010})}\BibitemShut {NoStop}%
\bibitem [{\citenamefont {Bregulla}\ \emph {et~al.}(2016)\citenamefont
  {Bregulla}, \citenamefont {W{\"u}rger}, \citenamefont {G{\"u}nther},
  \citenamefont {Mertig},\ and\ \citenamefont {Cichos}}]{bregulla2016thermo}%
  \BibitemOpen
  \bibfield  {author} {\bibinfo {author} {\bibfnamefont {A.~P.}\ \bibnamefont
  {Bregulla}}, \bibinfo {author} {\bibfnamefont {A.}~\bibnamefont
  {W{\"u}rger}}, \bibinfo {author} {\bibfnamefont {K.}~\bibnamefont
  {G{\"u}nther}}, \bibinfo {author} {\bibfnamefont {M.}~\bibnamefont {Mertig}},
  \ and\ \bibinfo {author} {\bibfnamefont {F.}~\bibnamefont {Cichos}},\
  }\bibfield  {title} {\enquote {\bibinfo {title} {Thermo-osmotic flow in thin
  films},}\ }\href@noop {} {\bibfield  {journal} {\bibinfo  {journal} {Phys.
  Rev. Lett.}\ }\textbf {\bibinfo {volume} {116}},\ \bibinfo {pages} {188303}
  (\bibinfo {year} {2016})}\BibitemShut {NoStop}%
\bibitem [{\citenamefont {McCann}\ \emph {et~al.}(1999)\citenamefont {McCann},
  \citenamefont {Dykman},\ and\ \citenamefont {Golding}}]{mccann1999thermally}%
  \BibitemOpen
  \bibfield  {author} {\bibinfo {author} {\bibfnamefont {L.~I.}\ \bibnamefont
  {McCann}}, \bibinfo {author} {\bibfnamefont {M.}~\bibnamefont {Dykman}}, \
  and\ \bibinfo {author} {\bibfnamefont {B.}~\bibnamefont {Golding}},\
  }\bibfield  {title} {\enquote {\bibinfo {title} {Thermally activated
  transitions in a bistable three-dimensional optical trap},}\ }\href@noop {}
  {\bibfield  {journal} {\bibinfo  {journal} {Nature}\ }\textbf {\bibinfo
  {volume} {402}},\ \bibinfo {pages} {785--787} (\bibinfo {year}
  {1999})}\BibitemShut {NoStop}%
\bibitem [{\citenamefont {Woodside}\ \emph {et~al.}(2006)\citenamefont
  {Woodside}, \citenamefont {Anthony}, \citenamefont {Behnke-Parks},
  \citenamefont {Larizadeh}, \citenamefont {Herschlag},\ and\ \citenamefont
  {Block}}]{woodside2006direct}%
  \BibitemOpen
  \bibfield  {author} {\bibinfo {author} {\bibfnamefont {M.~T.}\ \bibnamefont
  {Woodside}}, \bibinfo {author} {\bibfnamefont {P.~C.}\ \bibnamefont
  {Anthony}}, \bibinfo {author} {\bibfnamefont {W.~M.}\ \bibnamefont
  {Behnke-Parks}}, \bibinfo {author} {\bibfnamefont {K.}~\bibnamefont
  {Larizadeh}}, \bibinfo {author} {\bibfnamefont {D.}~\bibnamefont
  {Herschlag}}, \ and\ \bibinfo {author} {\bibfnamefont {S.~M.}\ \bibnamefont
  {Block}},\ }\bibfield  {title} {\enquote {\bibinfo {title} {Direct
  measurement of the full, sequence-dependent folding landscape of a nucleic
  acid},}\ }\href@noop {} {\bibfield  {journal} {\bibinfo  {journal} {Science}\
  }\textbf {\bibinfo {volume} {314}},\ \bibinfo {pages} {1001--1004} (\bibinfo
  {year} {2006})}\BibitemShut {NoStop}%
\bibitem [{\citenamefont {Volpe}\ and\ \citenamefont
  {Petrov}(2006)}]{volpe2006torque}%
  \BibitemOpen
  \bibfield  {author} {\bibinfo {author} {\bibfnamefont {G.}~\bibnamefont
  {Volpe}}\ and\ \bibinfo {author} {\bibfnamefont {D.}~\bibnamefont {Petrov}},\
  }\bibfield  {title} {\enquote {\bibinfo {title} {Torque detection using
  {B}rownian fluctuations},}\ }\href@noop {} {\bibfield  {journal} {\bibinfo
  {journal} {Phys. Rev. Lett.}\ }\textbf {\bibinfo {volume} {97}},\ \bibinfo
  {pages} {210603} (\bibinfo {year} {2006})}\BibitemShut {NoStop}%
\bibitem [{\citenamefont {Volpe}\ \emph {et~al.}(2007)\citenamefont {Volpe},
  \citenamefont {Volpe},\ and\ \citenamefont {Petrov}}]{volpe2007brownian}%
  \BibitemOpen
  \bibfield  {author} {\bibinfo {author} {\bibfnamefont {G.}~\bibnamefont
  {Volpe}}, \bibinfo {author} {\bibfnamefont {G.}~\bibnamefont {Volpe}}, \ and\
  \bibinfo {author} {\bibfnamefont {D.}~\bibnamefont {Petrov}},\ }\bibfield
  {title} {\enquote {\bibinfo {title} {Brownian motion in a nonhomogeneous
  force field and photonic force microscope},}\ }\href@noop {} {\bibfield
  {journal} {\bibinfo  {journal} {Phys. Rev. E}\ }\textbf {\bibinfo {volume}
  {76}},\ \bibinfo {pages} {061118} (\bibinfo {year} {2007})}\BibitemShut
  {NoStop}%
\bibitem [{\citenamefont {Blickle}\ \emph {et~al.}(2007)\citenamefont
  {Blickle}, \citenamefont {Speck}, \citenamefont {Lutz}, \citenamefont
  {Seifert},\ and\ \citenamefont {Bechinger}}]{blickle2007einstein}%
  \BibitemOpen
  \bibfield  {author} {\bibinfo {author} {\bibfnamefont {V.}~\bibnamefont
  {Blickle}}, \bibinfo {author} {\bibfnamefont {T.}~\bibnamefont {Speck}},
  \bibinfo {author} {\bibfnamefont {C.}~\bibnamefont {Lutz}}, \bibinfo {author}
  {\bibfnamefont {U.}~\bibnamefont {Seifert}}, \ and\ \bibinfo {author}
  {\bibfnamefont {C.}~\bibnamefont {Bechinger}},\ }\bibfield  {title} {\enquote
  {\bibinfo {title} {Einstein relation generalized to nonequilibrium},}\
  }\href@noop {} {\bibfield  {journal} {\bibinfo  {journal} {Phys. Rev. Lett.}\
  }\textbf {\bibinfo {volume} {98}},\ \bibinfo {pages} {210601} (\bibinfo
  {year} {2007})}\BibitemShut {NoStop}%
\bibitem [{\citenamefont {Gomez-Solano}\ \emph {et~al.}(2009)\citenamefont
  {Gomez-Solano}, \citenamefont {Petrosyan}, \citenamefont {Ciliberto},
  \citenamefont {Chetrite},\ and\ \citenamefont
  {Gaw\c{e}dzki}}]{gomez2009experimental}%
  \BibitemOpen
  \bibfield  {author} {\bibinfo {author} {\bibfnamefont {J.~R.}\ \bibnamefont
  {Gomez-Solano}}, \bibinfo {author} {\bibfnamefont {A.}~\bibnamefont
  {Petrosyan}}, \bibinfo {author} {\bibfnamefont {S.}~\bibnamefont
  {Ciliberto}}, \bibinfo {author} {\bibfnamefont {R.}~\bibnamefont {Chetrite}},
  \ and\ \bibinfo {author} {\bibfnamefont {K.}~\bibnamefont {Gaw\c{e}dzki}},\
  }\bibfield  {title} {\enquote {\bibinfo {title} {Experimental verification of
  a modified fluctuation-dissipation relation for a micron-sized particle in a
  nonequilibrium steady state},}\ }\href@noop {} {\bibfield  {journal}
  {\bibinfo  {journal} {Phys. Rev. Lett.}\ }\textbf {\bibinfo {volume} {103}},\
  \bibinfo {pages} {040601} (\bibinfo {year} {2009})}\BibitemShut {NoStop}%
\bibitem [{\citenamefont {Montiel}\ \emph {et~al.}(2006)\citenamefont
  {Montiel}, \citenamefont {Cang},\ and\ \citenamefont
  {Yang}}]{montiel2006quantitative}%
  \BibitemOpen
  \bibfield  {author} {\bibinfo {author} {\bibfnamefont {D.}~\bibnamefont
  {Montiel}}, \bibinfo {author} {\bibfnamefont {H.}~\bibnamefont {Cang}}, \
  and\ \bibinfo {author} {\bibfnamefont {H.}~\bibnamefont {Yang}},\ }\bibfield
  {title} {\enquote {\bibinfo {title} {{Quantitative characterization of
  changes in dynamical behavior for single-particle tracking studies.}}}\
  }\href {\doibase 10.1021/jp062024j} {\bibfield  {journal} {\bibinfo
  {journal} {J. Phys. Chem. B}\ }\textbf {\bibinfo {volume} {110}},\ \bibinfo
  {pages} {19763--70} (\bibinfo {year} {2006})}\BibitemShut {NoStop}%
\bibitem [{\citenamefont {Hummer}\ and\ \citenamefont
  {Szabo}(2010)}]{hummer2010free}%
  \BibitemOpen
  \bibfield  {author} {\bibinfo {author} {\bibfnamefont {G.}~\bibnamefont
  {Hummer}}\ and\ \bibinfo {author} {\bibfnamefont {A.}~\bibnamefont {Szabo}},\
  }\bibfield  {title} {\enquote {\bibinfo {title} {Free energy profiles from
  single-molecule pulling experiments},}\ }\href@noop {} {\bibfield  {journal}
  {\bibinfo  {journal} {PNAS}\ }\textbf {\bibinfo {volume} {107}},\ \bibinfo
  {pages} {21441--21446} (\bibinfo {year} {2010})}\BibitemShut {NoStop}%
\end{thebibliography}%

\end{document}